\documentclass[aps,showpacs,amssymb,superscriptaddress,nofootinbib]
{revtex4}

\usepackage{amssymb,amsmath}
\usepackage{amsfonts}
\usepackage{mathrsfs}

\renewcommand{\theequation}{\arabic{section}.\arabic{equation}}

\begin{document}


\title{Quantum Gowdy $T^3$ model: A unitary description}

\author{Alejandro Corichi}\email{corichi@matmor.unam.mx}
\affiliation{Instituto de Matem\'aticas, Universidad Nacional
Aut\'onoma de M\'exico\\ A. Postal 61-3, Morelia, Michoac\'an 58090,
Mexico.} \affiliation{Instituto de Ciencias Nucleares,
Universidad Nacional Aut\'onoma de M\'exico\\
A. Postal 70-543,
M\'exico D.F. 04510, Mexico.}
\author{ Jer\'onimo Cortez}\email{jacq@iem.cfmac.csic.es}
\affiliation{Instituto de Estructura de la Materia, CSIC, Serrano
121, 28006 Madrid, Spain.}
\author{Guillermo A. Mena Marug\'an}\email{mena@iem.cfmac.csic.es}
\affiliation{Instituto de Estructura de la Materia, CSIC, Serrano
121, 28006 Madrid, Spain.}

\begin{abstract}
The quantization of the family of linearly polarized Gowdy $T^{3}$
spacetimes is discussed in detail, starting with a canonical
analysis in which the true degrees of freedom are described by
a scalar field that satisfies a Klein-Gordon type equation in a
fiducial time dependent background. A time dependent canonical
transformation, which amounts to a change of the basic (scalar)
field of the model, brings the system to a description in terms of
a Klein-Gordon equation on a background that is now static,
although subject to a time dependent potential. The
system is quantized by means of a natural choice of annihilation and
creation operators. The quantum time evolution is considered and
shown to be unitary, allowing both the Schr\"odinger and Heisenberg
pictures to be consistently constructed. This has to be contrasted
with previous treatments for which time evolution failed to be
implementable as a unitary transformation. Possible implications for
both canonical quantum gravity and quantum field theory in curved
spacetime are commented.
\end{abstract}

\pacs{04.60.Ds, 04.62.+v, 04.60.Kz, 98.80.Qc}

\maketitle

\section{Introduction}

In the search for a quantum theory of gravity within the
canonical approach, it has always been useful to analyze symmetry
reduced models. On the one hand, this allows to discuss with
specific examples conceptual and technical issues that arise when
trying to conciliate gravity and quantum mechanics. On the other
hand, these reduced models are usually of physical relevance in
cosmology or in astrophysical situations. The most studied examples
are mini-superspaces \cite{misner1}, where the infinite dimensional
system is reduced by symmetry considerations to a model with a
finite number of degrees of freedom. A more interesting and
far-reaching class of reduced models, where the resulting system is
still a field theory with an infinite number of degrees of freedom
like general relativity, are known as midi-superspaces
\cite{torre}.

The simplest of all inhomogeneous midi-superspaces in pure
general relativity with spatially closed spatial sections and
cosmological solutions (expanding from a big-bang singularity) is
the linearly polarized Gowdy $T^3$ model \cite{gowdy}. This explains
the considerable attention that has been paid during the last thirty
years to the problem of quantizing this model
\cite{varios,berger1,berger,guillermo-gowdy,husain-smolin,
pierri,ccq-t3,torre-prd}.
After the first preliminary attempts to construct a quantization and
obtain physical predictions for the Gowdy $T^3$ cosmologies
employing conventional (but not always rigorously implemented)
canonical methods in quantum cosmology \cite{varios,berger1,berger},
the problem was revisited using Ashtekar variables in the context of
a nonperturbative quantization \cite{guillermo-gowdy,husain-smolin}.
Nonetheless, it is only recently that true progress has been
achieved in the task of introducing a consistent quantization, at
least for the (sub-)model with linear polarization \cite{pierri} for
which the two spacelike Killing vector fields of the system are
hypersurface orthogonal.

The quantization proposed in Ref. \cite{pierri} for the linearly
polarized Gowdy $T^3$ model is based on the equivalence that exists
between the set of solutions for its spacetime metric and the
classical solutions for a scalar field coupled to gravity in $2+1$
dimensions, defined in a manifold whose topology is
$\mathbb{R}^+\times T^2$. In more detail, after a suitable (partial)
gauge fixing of the Gowdy $T^3$ model, which includes the choice of
an internal time, the linearly polarized Gowdy $T^{3}$ spacetimes
are described (modulo a remaining global constraint) by a ``point
particle" degree of freedom and by a field $\phi$ that is subject to
the same equation of motion as a massless, rotationally symmetric,
free scalar field that propagates in a fictitious two-dimensional
expanding torus. The quantization of the local degrees of freedom of
the Gowdy model can hence be confronted by constructing a quantum
theory for this scalar field. The quantum Gowdy $T^{3}$ model is
defined by introducing a representation for the field $\phi$ on a
fiducial Fock space and imposing on it the constraint that remains
on the system as an operator condition, in order to finally obtain
the Hilbert space of physical states.

However, there is an important drawback to the quantization
presented in Ref.~\cite{pierri}. It can be proved that the quantum
evolution admits no implementation as a unitary transformation.
Moreover, this negative result applies to the implementation both on
the kinematical Hilbert space \cite{ccq-t3} and on the physical
Hilbert space of the model \cite{torre-prd}. To make things even
more subtle, it turns out that the dynamics can be approximated as
close as one wants in terms of unitary transformations \cite{jg}
but, nonetheless, the true evolution cannot be represented by a
unitary operator. Owing to this failure of unitarity, we do not have
at our disposal a Schr\"odinger picture with an evolution that
conserves the standard notion of probability \cite{torre-prd,jacob}.
In a more pessimistic note, one might argue that such nonunitarity
poses serious problems for a proper physical description of the
system \cite{ccq-t3}. A careful analysis of this issue would require
further discussion about the existence of the Heisenberg picture
when the Schr\"odinger one is not available \cite{heis-schr,SandG},
clarifying its physical validity and elucidating whether one should
or not abandon the concept of unitary evolution. We shall not pursue
this avenue here but, considering the Gowdy cosmology as a
particular arena in which one is addressing the issue of unitarity
in cosmology, we will rather show that the problems with the quantum
evolution can be solved, at least in this case, by adopting a
different quantization for the model.

In the study of quantum cosmological models, a fundamental issue is
the so-called problem of time. In general relativity there are no
preferred foliations in spacetime and the dynamical evolution should
consider all possible spacelike foliations. This is one of the main
features of diffeomorphism invariance. Furthermore, in cases with
compact Cauchy surfaces, dynamical evolution is pure gauge since
there is no true Hamiltonian. Then, any interpretation of time
evolution is normally obtained via a deparametrization which, in
Hamiltonian language, is achieved by fixing the time gauge. Thus,
the dynamics to be considered in these quantum cosmological models
concerns the evolution of quantum states between Cauchy surfaces,
defined by the particular choice of time gauge adopted. Different
choices of time may lead to inequivalent quantizations. In the
linearly polarized Gowdy $T^3$ model the system is partially gauge
fixed at the classical level and in particular a time function $t$
is chosen and interpreted as the time that defines ``evolution". The
surfaces of constant $t$ for the quantum gravity model turn out to
be Cauchy surfaces of the quantum scalar field in a fiducial
background equipped with a foliation of preferred surfaces.
Furthermore, even if there is no preferred time in the fundamental
description of quantum gravity, as well as in the cosmological Gowdy
models, and one would only expect a genuine notion of time to arise
in a certain semiclassical regime, the introduction of a
deparametrization allows to introduce a family of true observables,
the so called {\it evolving constants of motion}
\cite{rovelli1} that can be associated with quantities ``living at
time $t$''. A clean construction and interpretation of {these}
observables turn out to be possible in our case, since both a
Schr\"odinger and a Heisenberg pictures will be shown to exist.

This work has several aims. First, as we have just commented, we
will prove that it {\it is} indeed possible to achieve a unitary
quantum dynamics in the linearly polarized Gowdy cosmology.
Therefore, no fundamental obstruction exists to the standard
probabilistic interpretation of quantum physics in this
inhomogeneous cosmological framework. An outline of this result
was presented in Ref. \cite{ccmm}. The second aim of the present
paper is to systematically explore the canonical structure of the
Gowdy model. As we will show through a detailed analysis of the
implementation and consequences of a canonical transformation on
phase space, one can arrive at a suitable field-parametrization of
the spacetime metric of the model (i.e., to adopt an adequate
choice of basic field) which allows of a fully consistent
quantization. In addition, we also want to discuss some relevant
physical phenomena that occur in the model, e.g. the production of
``particles'' by the vacuum of the cosmological system and the
recovery of a time-translation invariance in the asymptotic region
of infinite large times.

The rest of the paper is organized as follows. Sec. \ref{sec:2}
reviews the standard formulation of the Gowdy $T^3$ model along the
lines put forward in Refs. \cite{pierri,ccq-t3,torre-prd,jg}
(including the realization of its nonunitary character). In Sec.
\ref{sec:3} we perform a time dependent canonical transformation and
find the corresponding new Hamiltonian for the model. The system is
then recast in Sec. \ref{sec:4} as a scalar field in a static
background with a potential. In that section we also find the
general classical solution for this scalar field and the finite
symplectic transformation associated with the time evolution. This
provides the starting point for Sec. \ref{sec:5}, where the quantum
representation is defined and the Fock space is constructed. The
issue of the unitarity of the evolution is analyzed in Sec.
\ref{sec:6}. It is shown that the symplectic transformations that
define finite time evolution can indeed be implemented in a unitary
way. Sec. \ref{sec:7} studies the properties of the quantum theory,
in particular its relation with previous work and the issue of
``particle production''. Finally, we present our conclusions and
some further discussion in Sec. \ref{sec:8}. An appendix is added,
where we show that the description of the model adopted in the main
text corresponds to the same gauge fixing that had been imposed in
previous treatments of the Gowdy $T^3$ cosmologies
\cite{pierri,torre-prd}.

\section{The polarized Gowdy $T^3$ model}
\label{sec:2}
\setcounter{equation}{0}

Let us briefly review the symmetry reduced model employed
in Ref. \cite{pierri} to introduce a quantum theory for the
linearly polarized Gowdy $T^3$ cosmologies. These cosmologies are
described by vacuum spacetimes with the spatial topology of a
three-torus that possess two commuting, axial, and hypersurface
orthogonal Killing vector fields \cite{gowdy}. They provide the
simplest of all possible cosmological models with compact spatial
sections and local degrees of freedom. After a gauge fixing
procedure, in which all the gauge degrees of freedom are removed
except for a homogenous one \cite{jg,guillermo-varios}, the metric
of the symmetry reduced model can be expressed in the form
\begin{eqnarray} ds^2&=& e^{\gamma} e^{ - \phi/\sqrt{p}}
\left(-dt^2+ d\theta^2\right)+e^{-\phi/\sqrt{p} } t^2 p^2d\sigma^2 +
e^{\phi/\sqrt{p}}d\delta^2,\label{metri}\\
\gamma&=& -\frac{\bar{Q}}{2\pi p}-\sum_{n=-\infty,n\neq
0}^{\infty}\frac{i}{2\pi n p}\oint d\bar{\theta}
e^{in(\theta-\bar{\theta})}P_{\phi}\phi^{\prime}+\frac{1}{4\pi
p}\oint d\bar{\theta}\left[P_{\phi}\phi+
P_{\phi}^2+t^2(\phi^{\prime})^2\right].\label{gamma}
\end{eqnarray}
Here, $t>0$ is a positive time coordinate and the angular
coordinates $\theta$, $\sigma$, and $\delta$ belong to $S^1$. The
two Killing vector fields are given by $\partial_{\sigma}$ and
$\partial_{\delta}$, so that the metric functions are independent of
these angles. The function $\phi$ and its canonical momentum
$P_{\phi}$ may depend on both $t$ and $\theta$. Therefore, they
describe a fieldlike degree of freedom of the metric. All cyclic
integrals in Eq. (\ref{gamma}) are performed on the corresponding
angular dependence and the prime stands for the derivative with
respect to $\theta$. On the other hand, $\bar{Q}$ and $p$ are
homogenous variables \cite{jg}. Moreover, $p$ is a constant of
motion that we impose to be strictly positive. Actually, spacetimes
with $p<0$ can be related with those with $p>0$ by means of a time
reversal, whereas spacetimes with vanishing $p$ can be consistently
removed (since the considered sector of phase space is dynamically
invariant). To avoid dealing with the positivity restriction, we
introduce the definition
\begin{equation}\label{P} \bar{P}:=\ln{p}.\end{equation}
The real constant of motion $\bar{P}$ provides the momentum
canonically conjugate to $\bar{Q}$. So, the pair $(\bar{Q},\bar{P})$
describes a ``point particle'' degree of freedom.

To be completely rigorous, the metric (\ref{metri}) should include a
nonvanishing $\theta$-component of the shift vector equal to an
arbitrary function of time \cite{jg}. Nonetheless, since this kind
of shift can always be absorbed by means of a redefinition of the
angular coordinate $\theta$, we have obviated it. This freedom in
the choice of shift appears because the gauge has not been totally
fixed. There is still a global constraint remaining on the system,
coming from the homogeneous part of the $\theta$-momentum
constraint:
\begin{equation} \label{glob-const}
C_{0}:=\frac{1}{\sqrt{2\pi}}\oint d\theta P_{\phi}\phi'=0 .
\end{equation}

Starting with the Einstein-Hilbert action of general relativity with
$4G/\pi=c=1$ ($G$ and $c$ being Newton's constant and the speed of
light, respectively) and after the above reduction process, one
arrives at the following action for the model (modulo the constraint
$C_0=0$ and spurious surface terms):
\begin{equation} S_{r}=\int_{t_{i}}^{t_{f}}dt \left(\bar{P}
\dot{\bar{Q}}+\oint d \theta\left[P_{\phi}\dot{\phi}-{{\cal H}}_{r}
\right] \right),\qquad {{\cal H}}_{r} =
\frac{1}{2t}\left[P_{\phi}^{2}+
t^2(\phi^{\prime})^{2}\right],\label{redac}
\end{equation}
where the dot denotes the derivative with respect to $t$. The
reduced Hamiltonian is
\begin{equation}\label{reduced-ham} H_r=\oint d\theta {\cal H}_{r}.
\end{equation} This Hamiltonian does not depend on the
``point particle'' degrees of freedom. Hence, $\bar{Q}$ and
$\bar{P}$ are constants of motion and a nontrivial evolution occurs
only in the field sector of the system. In addition note that, were
not for the explicit time dependence of ${\cal H}_r$, the reduced
Hamiltonian would be that of a massless scalar field with axial
symmetry in a static background.

The (reduced) phase space of the system ${\Gamma}_{r}$ can be
decomposed as a direct sum, ${\Gamma}_{r}=\Gamma_{0}\oplus
\tilde{\Gamma}$, where $\Gamma_{0}$ and $\tilde{\Gamma}$ contain the
``point particle" and the fieldlike degrees of freedom,
respectively. They admit as coordinates the canonical pairs
$(\bar{Q},\bar{P})$ and $\big(\phi, P_{\phi}\big)$. Owing to the
presence of the global constraint $C_0=0$, the space of physical
states does not really correspond to $\Gamma_{r}$, but rather to a
submanifold of it. However, since this submanifold is nonlinear, the
reduction by the constraint is postponed to the quantum theory,
where it is imposed as an operator condition on the (kinematical)
quantum states.

The Hamiltonian equations derived from $H_r$ require that the field
$\phi$ and its momentum satisfy \begin{equation}
\label{canonical-f-eq} \dot{\phi}=\frac{P_{\phi}}{t} , \qquad
\dot{P}_{\phi}=t \phi^{\prime\prime} . \end{equation} Combining
them, one concludes that $\phi$ is subject to a wave equation:
\begin{equation} \label{eq-phi}
\ddot{\phi}+\frac{\dot{\phi}}{t}-\phi^{\prime\prime}=0 .
\end{equation} We will
call $\varphi$ any smooth solution to this equation. Let us define
$f_{n}(t,\theta):=\bar{f}_{n}(t)\exp[in\theta]$ for all $n\in
\mathbb{Z}$ with
\begin{equation} \bar{f}_{0}(t):=\frac{1-i\ln
t}{\sqrt{4\pi}}, \qquad \bar{f}_{n}(t):=\frac{H_{0}(|n|t)}{\sqrt{8}}
\quad {\rm if} \quad n\neq 0.\label{fbar}
\end{equation} Here, $H_{0}$ is the zeroth-order Hankel function of
the second kind \cite{abramowitz}. We can then express all solutions
$\varphi$ in the generic form \begin{equation}
\label{scalarfield-sol} \varphi(t,\theta)=\sum_{n=-\infty}^{\infty}
\left[A_{n}f_{n}(t,\theta)+A^{*}_{n}f^{*}_{n}(t,\theta)\right].
\end{equation}
The symbol $*$ represents complex conjugation, and the $A_{n}$'s are
complex constant coefficients. These coefficients must decrease
faster than the inverse of any polynomial in $n$ as $|n|\to\infty$
in order to guarantee the pointwise convergence of the series
(\ref{scalarfield-sol}).

From action (\ref{redac}), one obtains the following symplectic
structure on the space $\{\varphi\}$ of smooth solutions to Eq.
(\ref{eq-phi}):
\begin{equation}\label{symp}
\tilde{\Omega}(\varphi_{1},\varphi_{2})=\oint d\theta
\left[\varphi_{2} t\partial_{t}\varphi_{1}-\varphi_{1}
t\partial_{t}\varphi_{2}\right],
\end{equation}
where the integral is taken over any $t=$constant slice. The set of
mode solutions $\{f_{n}(t,\theta),f^*_n(t,\theta)\}$ (with $n\in
\mathbb{Z}$) is complete and ``orthonormal" in the product
$(f_{l},f_{n})_{\varphi}=-i\tilde{\Omega}(f_{l}^{*},f_{n})$, in the
sense that $(f_{l},f_{n})_{\varphi}=\delta_{ln}$,
$(f^{*}_{l},f^{*}_{n})_{\varphi}=-\delta_{ln}$, and
$(f_{l},f^{*}_{n})_{\varphi}=0$. As a consequence, it is not
difficult to check that the complex conjugate constants
$\{A_{n},A^*_{n}\}$ behave as pairs of annihilationlike and
creationlike variables under this symplectic structure.

Returning to Eq. (\ref{eq-phi}), it is worth pointing out that it is
formally identical to the Klein-Gordon equation of a free massless
scalar field propagating in a fictitious three-dimensional
background $({\cal{M}}\simeq \mathbb{R}^{+}\times T^{2} ,g^{(B)})$,
where the background metric is $g^{(B)}_{ab}=-dt_adt_b+d\theta_a
d\theta_b +t^{2}d\sigma_ad\sigma_b$, again with $t\in
\mathbb{R}^{+}$ and $\theta ,\sigma \in S^1$. We can then identify
$\tilde{\Gamma}$ (the field part of phase space) with the canonical
phase space of the scalar field in that background, while the space
of smooth solutions can be considered as the covariant phase space
of such a Klein-Gordon field. That is, $\phi$ and $P_{\phi}$ can be
viewed as the configuration and momentum on the constant-time
section $\Sigma_{t}\simeq T^{2}$ of the scalar field $\varphi$
propagating in $({\cal M},g^{(B)})$.

A consequence of this equivalence between the gauge fixed Gowdy
model and a Klein-Gordon field is that, in this description, the
problem of quantization of the local degrees of freedom reduces to
the construction of a quantum theory for the axially symmetric
massless scalar field $\varphi$ in the background $({\cal
M},g^{(B)})$, which is an expanding torus. Employing this fact, the
quantum Gowdy $T^{3}$ model was defined in Ref. \cite{pierri} by
introducing a Fock representation for $\varphi$ and, on the fiducial
Fock space obtained in this way, imposing the global constraint
(\ref{glob-const}) as an operator condition in order to get the
physical Hilbert space. More precisely, taking into account the
field decomposition (\ref{scalarfield-sol}) [and remembering
definition (\ref{fbar})], the symplectic vector space
$\tilde{S}:=(\tilde{\Omega},\{\varphi\})$ can be endowed with the
$\tilde{\Omega}$-compatible complex structure
$\tilde{J}:\tilde{S}\to \tilde{S}$:
\begin{equation} \label{pierris-cs}
\tilde{J}\big[\bar{f}_{n}(t)\big]=i\bar{f}_{n}(t), \qquad
\tilde{J}\big[\bar{f}^{*}_{n}(t)\big]=-i\bar{f}^{*}_{n}(t).
\end{equation} Using this complex structure, it is
straightforward to construct from $\tilde{S}$ the ``one-particle"
Hilbert space of the theory, ${\tilde{\cal{H}}}$. This space allows
to define in turn the (symmetric) Fock space
${\cal{F}}(\tilde{\cal{H}})$ on which one can introduce the formal
field operator, expressed in terms of annihilation and creation
operators that correspond to the positive and negative frequency
parts determined by the complex structure $\tilde{J}$. Finally, one
specifies the explicit operator that represents the constraint
$C_0$ quantum mechanically on ${\cal{F}}(\tilde{\cal{H}})$. The
physical Hilbert space $\tilde{\cal{F}}_{\rm{phys}}$ is supplied by
the kernel of this operator.

In the quantization sketched above, however, it is known that the
dynamics dictated by the Hamiltonian $H_r$ [see Eqs. (\ref{redac})
and (\ref{reduced-ham})] cannot be implemented as a unitary
transformation. This result applies not only to the kinematical Fock
space ${\cal{F}}(\tilde{\cal{H}})$ \cite{ccq-t3}, but also to
the physical Hilbert space $\tilde{\cal{F}}_{\rm{phys}}$
\cite{torre-prd}. The antilinear part of the Bogoliubov
transformation that implements the classical dynamics in the quantum
theory, providing the relation between annihilation and creation
operators at different times $t_{0}$ and $t_{f}$, has a single
contribution for each of the field modes. The corresponding
Bogoliubov coefficient can be deduced employing the
``orthonormalization'' of the set
$\{f_{n}(t,\theta),f^*_n(t,\theta)\}$ (with $n\in \mathbb{Z}$) and
formula (\ref{symp}). For the nonzero modes ($n\neq 0$), the
coefficient is given by \cite{ccq-t3,torre-prd,jg}:
\begin{eqnarray}  D_{n}(t_{f},t_{0})&=&i 2\pi
\left[\bar{f}^{*}_{n}(t_{0})t_{f}\partial_t\bar{f}^{*}_{n}(t_{f})
-\bar{f}^*_{n}(t_{f})t_{0}\partial_t\bar{f}^{*}_{n}(t_{0})\right]
\label{Dsymp}\\ &=&
\frac{i\pi|n|}{4}\left[H^*_{0}(|n|t_{f})t_{0}H^{*}_{1}(|n|t_{0})
-H^{*}_{0}(|n|t_{0})t_{f}H^{*}_{1}(|n|t_{f})\right],
\label{antilin-sol-space}
\end{eqnarray} where $H_{1}$ is the first-order Hankel function of
the second kind \cite{abramowitz}. Since the sequence
$\{D_{n}(t_f,t_0)\}$ is not square summable for generic positive
times $t_{f}\neq t_{0}$ \cite{ccq-t3}, unitary implementability
is impossible \cite{shale,honegger}.

Actually, the failure of square summability can be easily derived by
making use of Hankel's asymptotic expansions for the functions $H_0$
and $H_1$ \cite{abramowitz}. Up to corrections of relative order
$1/t$ in these expansions,
\begin{equation}\label{hank}
H_{0}(|nt|)\approx \sqrt{\frac{2}{\pi
|n|t}}e^{-i|n|t}e^{i\pi/4},\qquad H_{1}(|nt|)\approx
i\sqrt{\frac{2}{\pi |n|t}}e^{-i|n|t}e^{i\pi/4}.\end{equation} From
these formulas and Eq. (\ref{antilin-sol-space}), one obtains at
leading order
\begin{equation} \label{approx-anti-old}
|D_{n}(t_{f},t_{0})|^{2}\approx
\frac{(t_{f}-t_{0})^{2}}{4t_{f}t_{0}}. \end{equation} Thus, for
asymptotically large values of $|n|$, $|D_{n}(t_{f},t_{0})|^{2}$
differs from zero if $t_{f}$ and $t_{0}$ do not coincide, so
that the sequence $\{D_{n}(t_f,t_0)\}$ is not square summable.

\section{New field-parametrization}
\label{sec:3}
\setcounter{equation}{0}

The above discussion shows that the failure of unitary
implementability of the dynamics can be blamed on the inadequate
asymptotic behavior of the Bogoliubov coefficient
$D_{n}(t_{f},t_{0})$. Furthermore, a close look at Eq. (\ref{Dsymp})
reveals that this problematic behavior can be traced back to the
appearance of the factor $t$ in the symplectic structure
(\ref{symp}). The obvious way to get rid of this unwanted factor in
the symplectic (two-)form is to absorb its square root, $\sqrt{t}$,
in $\varphi$. Notice that such rescaled solutions are nothing but
the classical smooth solutions for the rescaled field
$\sqrt{t}\phi$. The proposed change results then in a rescaling of
the complete set of mode solutions
$\{f_n(t,\theta),f_n^*(t,\theta)\}$ ($n\in \mathbb{Z}$) into the set
$\{g_{n}(t,\theta),g_{n}^*(t,\theta)\}:=\{\sqrt{t}
f_n(t,\theta),\sqrt{t}f_n^*(t,\theta)\}$, which is again a complete
set of solutions, but now for the classical equation of motion
satisfied by the rescaled field.

Actually, the asymptotic behavior of the nonzero mode solutions
$f_n(t,\theta)$ strongly suggests the proposed rescaling also from
the following related perspective. Using Hankel's expansions for the
nonzero modes we see that, both for asymptotically large wave
numbers $|n|$ and for asymptotically large times $t$,
\begin{equation}\label{asympmode} f_{n}(t,\theta)\approx
\frac{e^{i\pi/4}}{\sqrt{4\pi|n|t}} e^{-i(|n|t-n\theta)}.
\end{equation}
Therefore, in these asymptotic regimes (and modulo negligible
corrections compared with the unity) the functions
$\sqrt{t}f_{n}(t,\theta)$ (with $n\neq 0$) behave like the standard
nonzero mode solutions corresponding to an axially symmetric, free,
and massless scalar field propagating in the static background
$({\cal{M}},g^{(S)})$, with ${\cal{M}}\simeq \mathbb{R}^{+}\times
T^{2}$ and
\begin{equation}\label{stabac}
g^{(S)}_{ab}=-dt_adt_b+d\theta_a d\theta_b +d\sigma_a
d\sigma_b.\end{equation} Thus, the change from $\varphi$ to
$\sqrt{t}\varphi$ should provide us with a description that
corresponds asymptotically to a massless scalar field in Minkowski
spacetime (except for the topology). We recall that time evolution
between any two flat Cauchy surfaces is unitarily implementable for
the latter system.

With these motivations, we will now proceed to reformulate the
linearly polarized Gowdy model by considering as our new
covariant phase space the set $\{\sqrt{t}\varphi\}$. We will also
see that, for asymptotically large values of $t$, the dynamics of
the (nonzero modes of the) system is indeed dictated by the
Hamiltonian of a collection of harmonic oscillators with frequencies
$\omega_{n}=|n|$.

Let us hence start by multiplying the field $\phi$ by the factor
$\sqrt{t}$ and completing this rescaling into a time dependent
canonical transformation in the most straightforward
way\footnote{This kind of transformation was already considered in
Ref. \cite{berger1} although in a different and restricted context,
namely the study of the WKB regime.}, namely
\begin{equation}
\label{cano1-transf} \chi:=\sqrt{t} \phi, \qquad
P_{\chi}:=\frac{1}{\sqrt{t}} P_{\phi}.
\end{equation}
Obviously, the canonical variables $(\bar{Q},\bar{P})$ for the
``point particle'' degrees of freedom need not be changed. The above
transformation allows us to rewrite the Hamiltonian
(\ref{reduced-ham}) as that of an axially symmetric free field,
\begin{equation}\label{Hchifield}
H_{r}=\frac{1}{2}\oint d\theta
\left[P_{\chi}^{2}+(\chi^{\prime})^2\right].
\end{equation}
On the other hand, given the periodicity of the system in $\theta$,
we can expand $\chi$ and $P_{\chi}$ in Fourier series,
\begin{equation}\label{fourier}
\chi=\sum_{n=-\infty}^{\infty}\chi_{(n)}
\frac{e^{in\theta}}{\sqrt{2\pi}},\qquad
P_{\chi}=\sum_{n=-\infty}^{\infty}P_{\chi}^{(n)}
\frac{e^{in\theta}}{\sqrt{2\pi}}.
\end{equation}
It is not difficult to check that these (implicitly time dependent)
Fourier coefficients form canonical pairs, with $P_{\chi}^{(-n)}$
being the momentum conjugate to $\chi_{(n)}$. In terms of them, the
Hamiltonian $H_r$ adopts the expression
\begin{equation}\label{Hchi} H_{r}=\frac{1}{2}
\sum_{n=-\infty}^{\infty} \left[P_{\chi}^{(n)}P_{\chi}^{(-n)}+n^{2}
\chi_{(n)}\chi_{(-n)}\right].\end{equation} Then, $H_r$ can be
equivalently interpreted as describing a free particle (the zero
mode $n=0$) and a combination of harmonic oscillators with
frequencies equal to $|n|$ [two for each value of $n\neq 0$,
corresponding to the real and imaginary parts of $\chi_{(n)}$
and $P_{\chi}^{(-n)}$]. Our mode decomposition leads in this way to
a natural choice of annihilationlike variables (up to trivial
linear combinations),
\begin{equation}\label{annicrea}
a_n=\frac{|n|
\chi_{(n)}+iP_{\chi}^{(n)}}{\sqrt{2|n|}}\end{equation} for all
$n\neq 0$, with creationlike variables obtained by complex
conjugation.

However, since the change (\ref{cano1-transf}) is a time dependent
canonical transformation, the dynamical evolution in the new
field-parametrization is not generated by $H_r$ anymore. Instead,
the new reduced Hamiltonian of the system is
$\tilde{H}_{r}=H_{r}+\partial_{t}\tilde{F}$, where the partial
derivative refers only to the explicit time dependence and
$\tilde{F}$ is a generating functional for the canonical
transformation. For instance, we can choose $\tilde{F}=-\oint
P_{\phi}\chi/\sqrt{t}$. Then $\partial_{t}\tilde{F}=\oint
P_{\chi}\chi/(2t)$ and
\begin{equation} \label{bad} \tilde{H}_{r}=\frac{1}{2}\oint d\theta
\left[P_{\chi}^{2}+(\chi^{\prime})^{2}+\frac{P_{\chi}\chi}{t}
\right].
\end{equation}
One may decompose $\tilde{H}_r$ in Fourier modes in a similar way to
what we did with $H_r$ and express the result in terms of the
annihilationlike and creationlike variables introduced in
Eq. (\ref{annicrea}). Modulo a point particle, the new Hamiltonian
corresponds to an infinite number of standard harmonic oscillators
in the limit $t\rightarrow \infty$, as expected from Eq.
(\ref{bad}).

In spite of this good feature, there is a serious problem that
makes us disregard $\tilde{H}_r$ as a suitable Hamiltonian for
the Gowdy model. In the Fock representation that (for the
nonzero modes of the system) determines the choice (\ref{annicrea})
of annihilationlike and creationlike variables, one can easily
check that the vacuum does not belong to the domain of the operator
counterpart of $\tilde{H}_r$: the action of the operator on the
vacuum has infinite norm. As a consequence, it follows that such an
operator cannot be defined on the dense subspace of the
(kinematical) Hilbert space formed by the states with a finite
number of particles (indeed, none of these states has a normalizable
image). Actually, this problem appears owing to the presence of the
cross term $\oint P_{\chi}\chi/(2t)$ in the Hamiltonian.
Nonetheless, this term is negligible in the limit of infinite large
times and is not required to arrive at the desired asymptotic
behavior for the system.

In fact, the commented cross term can be eliminated by including a
linear contribution of the field $\chi$ in its canonical momentum.
As a result, the time dependent canonical transformation
(\ref{cano1-transf}) is replaced with the following one:
\begin{equation} \label{cano-transf}
\xi:=\sqrt{t} \phi, \qquad
P_{\xi}:=\frac{1}{\sqrt{t}}\left(P_{\phi}+\frac{\phi}{2}\right).
\end{equation}
This transformation is generated by the functional
\begin{equation}\label{generfunc2}
\bar{F}=-\oint d\theta
\left(\frac{P_{\phi}\xi}{\sqrt{t}}+\frac{\xi^{2}}{4t}\right).
\end{equation}
Then, the reduced action and Hamiltonian become
\begin{eqnarray}
\label{action-new-var} S_{r}&=&\int_{t_{i}}^{t_{f}}dt
\left(\bar{P}\dot{\bar{Q}}+ \oint d\theta \left[
P_{\xi}\dot{\xi} -\bar{{\cal H}}_{r}\right] \right), \\
\label{hamil-new-var} \bar{H}_{r}&=&\oint d\theta \;\bar{{\cal
H}}_r=\frac{1}{2}\oint d\theta
\left[P_{\xi}^{2}+(\xi^{\prime})^{2}+\frac{\xi^2}{4t^{2}} \right].
\end{eqnarray} We notice that the new reduced Hamiltonian is just
that of an axially symmetric Klein-Gordon field propagating in
fact in the fictitious static background $({\cal{M}}\approx
\mathbb{R}^+ \times T^{2} , g^{(S)})$ [see Eq. (\ref{stabac})],
though now subject to a time dependent potential that corresponds to
an effective mass equal to $1/(2t)$. Note nevertheless that this
potential vanishes in the limit of large times.

It is worth remarking that the time dependent canonical
transformation (\ref{cano-transf}), introduced to recast the
symmetry reduced model in terms of a new canonical set of variables
$(\xi,P_{\xi})$, amounts just to a field-reparametrization of the
spacetime metric of the linearly polarized Gowdy model. The
construction of the reduced model in terms of the canonical
pair $(\xi,P_{\xi})$ is completely parallel to that explained in
Ref. \cite{jg} for the variables $(\phi,P_{\phi})$ (which is
essentially the description considered in Refs.
\cite{berger1,pierri,ccq-t3,torre-prd}), the only difference being
the distinct parametrization of the metric. In all other
respects, the gauge fixing and reduction process is the same,
including the conditions imposed to fix almost entirely the gauge
freedom. We show this in detail in Appendix \ref{app-gt3-new}.

\section{Dynamics in the new description}
\label{sec:4}
\setcounter{equation}{0}

Varying action (\ref{action-new-var}) one recovers the result that
the ``point particle'' degrees of freedom $(\bar{Q},\bar{P})$ remain
constant in the evolution, whereas the dynamics in the field sector
is dictated by
\begin{equation} \label{xidyn} \dot{\xi}=P_{\xi}, \qquad
\dot{P}_{\xi}=\xi^{\prime\prime}-\frac{\xi}{4t^{2}}. \end{equation}
The field $\xi$ must then satisfy the second-order differential
equation
\begin{equation} \label{scalarf.new-var}
\ddot{\xi}-\xi^{\prime\prime}+\frac{\xi}{4t^{2}}=0.
\end{equation} From Eqs. (\ref{scalarfield-sol}) and
(\ref{cano-transf}) we have that all smooth solutions $\zeta$ to Eq.
(\ref{scalarf.new-var}) have the general form \begin{equation}
\label{scalarf-in-a-var} \zeta(t,\theta)=\sum_{n=-\infty}^{\infty}
\left[A_{n}g_{n}(t,\theta)+A_{n}^{*}g_{n}^{*}(t,\theta) \right].
\end{equation} We recall that
$g_{n}(t,\theta):=\sqrt{t}f_{n}(t,\theta)$. As anticipated at the
beginning of Sec. III, the set of mode solutions
$\{g_{n}(t,\theta),g_{n}^*(t,\theta)\}$ is complete.

On the other hand, the symplectic structure on the field sector of
the canonical phase space is
\begin{equation}\label{symcanoxi}
\Lambda([\xi_{1},P_{\xi_1}],[\xi_{2},P_{\xi_2}])=\oint d\theta
\left(\xi_{2} P_{\xi_1}-\xi_{1} P_{\xi_2}\right). \end{equation}
Taking into account the first of Eqs. (\ref{xidyn}), this leads to
the following symplectic structure on the space $\{\zeta\}$ of
smooth solutions:
 \begin{equation}\label{sympxi} \Omega(\zeta_{1},\zeta_{2}) =
  \oint d\theta\left(\zeta_{2}\partial_{t}\zeta_{1}-\zeta_{1}
\partial_{t}\zeta_{2}\right).\end{equation}
Comparing this with the symplectic structure (\ref{symp}) for the
``old'' field-parametrization of the Gowdy model, we see that the
problematic factor $t$ has indeed disappeared. In addition, the set
of mode solutions $\{g_{n}(t,\theta),g_{n}^*(t,\theta)\}$ is
``orthonormal" in the product
$(g_{l},g_{n})_{\zeta}=-i\Omega(g_{l}^{*},g_{n})$, as required for
consistency since the set $\{f_{n}(t,\theta),f_{n}^*(t,\theta)\}$ is
``orthonormal'' with respect to the corresponding inner product in
$(\tilde{\Omega},\{\varphi\})$.

Therefore, for the reformulated Gowdy model, the field sector of the
covariant phase space is the symplectic vector space
$S:=(\Omega,\{\zeta\})$, which can be coordinatized either by the
set $\{\zeta\}$ of smooth solutions to Eq. (\ref{scalarf.new-var})
or by the complex constants of motion
$(\mathsf{A}_{0},\{{\cal{A}}_{m}\})$, where
${\mathsf{A}}_{0}:=(A_{0},A_{0}^{*})$ and
${\cal{A}}_{m}:=(A_{m},A_{m}^{*} , A_{-m} , A_{-m}^{*})$ for all ${m
\in \mathbb{N} -\{0\}}$. Remember that these constants are of the
annihilation and creation type. Alternatively, one can consider the
canonical phase space of the model, whose field sector is the
symplectic vector space $\Gamma:=\big(\Lambda ,
\{(\xi,P_{\xi})\}\big)$, coordinatized by the configuration $\xi$
and momentum $P_{\xi}$ of the massive scalar field. Expanding in
Fourier series our canonical variables $\xi$ and $P_{\xi}$, as we
did for $\chi$ and $P_{\chi}$ in Eq. (\ref{fourier}), we can
equivalently adopt as coordinates the set of (complex) canonical
pairs $\{\xi_{(n)},P_{\xi}^{(-n)}\}$, with $n\in \mathbb{Z}$. We
emphasize that these Fourier coefficients depend implicitly on the
time coordinate $t$.

Analogously to our definition in Eq. (\ref{annicrea}), we now
introduce annihilationlike and creationlike variables for
the nonzero modes $n\neq 0$,
\begin{equation}\label{bns}
b_{n}=\frac{|n| \xi_{(n)}+iP_{\xi}^{(n)}}{\sqrt{2|n|}}, \qquad
b_{-n}^{*}=\frac{|n| \xi_{(n)}-iP_{\xi}^{(n)}}{\sqrt{2|n|}}.
\end{equation}
For convenience, we introduce a similar change of variables for
the zero mode, although the behavior of this mode corresponds to a
free particle in the limit of asymptotically large times, rather
than to an oscillator:
\begin{equation}\label{b0s}
b_{0}=\frac{\xi_{(0)}+iP_{\xi}^{(0)}}{\sqrt{2}}, \qquad b_{0}^{*}=
\frac{\xi_{(0)}-iP_{\xi}^{(0)}}{\sqrt{2}}.
\end{equation}
These transformations are canonical inasmuch as the pairs $b_{n}$
and $ib_{n}^{*}$ are canonically conjugate for all values of $n$.
Hence, we have that $\Gamma$ can be alternatively coordinatized by
the complex variables $({\mathsf{B}}_{0},\{{\cal{B}}_{m}\})$, where
${\mathsf{B}}_{0}:=(b_{0},b_{0}^{*})$ and ${\cal{B}}_{m}:=(b_{m} ,
b_{m}^{*}, b_{-m} , b_{-m}^{*})$ for all $m\in \mathbb{N}-\{0\}$.
The physical phase space consists then of those states
$({\mathsf{B}}_{0},\{{\cal{B}}_{m}\})$ in $\Gamma$ that satisfy the
constraint $C_0$, which can be expressed in the form
\begin{equation} \label{gconst-b-var}
C_{0}=\sum_{m=1}^{\infty}m
\left(b_{m}^{*}b_{m}-b_{-m}^{*}b_{-m}\right)=0 .
\end{equation}
Since this constraint defines a non-linear submanifold of $\Gamma$,
we leave the corresponding reduction to the quantum theory.

The dynamics on $\Gamma$ is dictated by the reduced Hamiltonian
(\ref{hamil-new-var}), which in terms of the new set of variables is
given by
\begin{equation} \label{hamil-in-b-coord}
\bar{H}_{r} = \sum_{m=0}^{\infty}
\left[\omega_{(m,t)}(b_{m}^{*}b_{m}+b^{*}_{-m}b_{-m}) +
\rho_{(m,t)}(b_{m}^{*}b_{-m}^{*}+b_{m}b_{-m})\right],
\end{equation}
where
\begin{eqnarray}\label{coeham0}
\omega_{(0,t)}&:=&\frac{1}{4}+
\frac{1}{16t^{2}},\qquad
\rho_{(0,t)}:=-\frac{1}{4}+\frac{1}{16t^{2}},
\\
\omega_{(m,t)}&:=&m+\frac{1}{8mt^{2}},\qquad
\rho_{(m,t)}:=\frac{1}{8mt^{2}},
\end{eqnarray}
for the zero and nonzero modes, respectively. In coordinates
$(\xi_{(0)},P^{(0)}_{\xi})$, the zero mode part of this Hamiltonian
$\bar{H}_{r}^{(0)}$ can be written
\begin{equation}\bar{H}_{r}^{(0)}:= \frac{
\big(P^{(0)}_{\xi}\big)^{2}}{2}+
\frac{\big(\xi_{(0)}\big)^{2}}{8t^{2}}.
\end{equation}
In agreement with our above comments, this Hamiltonian describes a
harmonic oscillator with unit mass and frequency $\omega=1/(2t)$
that behaves asymptotically like a free particle. On the other hand,
for the nonzero modes, we get $\lim_{t\to \infty}\omega_{(m,t)}=m$
and $\lim_{t\to \infty}\rho_{(m,t)}=0$. Thus, asymptotically, the
dynamics of the nonzero modes corresponds in fact to that of a
collection of harmonic oscillators with frequency $\omega_{n}=|n|$.
Besides, in contrast to the situation found in Sec. III with the
field $\chi$, the vacuum will be in the domain of the Hamiltonian
operator since $\rho_{(m,t)}$ turns out to be square summable.

The map from the covariant phase space $S$ to the canonical phase
space $\Gamma$ is given in the case of the zero mode by
${\mathsf{B}}_{0}(t)=W_{0}(t){\mathsf{A}}_{0}$ (treating
${\mathsf{B}}_{0}$ and ${\mathsf{A}}_{0}$ as row vectors), where
\begin{eqnarray}
\label{zero-modes-map} W_{0}(t) &=& \left(\begin{array}{cc} r_{0}(t)
& s_{0}(t)  \\ s_{0}^{*}(t) & r_{0}^{*}(t)\end{array} \right)
\end{eqnarray} with \begin{equation}\label{c-and-d-coeff0}
r_{0}(t):=\sqrt{\pi}g_{0}(t,\theta)\left(1+\frac{i}{2t}\right)
+\frac{1}{2\sqrt{t}},\qquad
s_{0}(t):=\sqrt{\pi}g^{*}_{0}(t,\theta)\left(1+\frac{i}{2t}
\right)-\frac{1}{2\sqrt{t}}.\end{equation}

For the remaining modes ($m\in \mathbb{N}-\{0\}$) the map is
${\cal{B}}_{m}(t)=W(x_{m}){\cal{A}}_{m}$, where we have defined
$x_{m}:=mt$, \begin{equation}\label{nonzer-modes-map} W(x_{m})=
\left(\begin{array}{cccc} c(x_{m}) & 0 & 0 & d(x_{m}) \\ 0 &
c^{*}(x_{m}) & d^{*}(x_{m}) & 0
\\ 0 & d(x_{m}) & c(x_{m}) & 0 \\ d^{*}(x_{m}) & 0 & 0 &
c^{*}(x_{m}) \end{array} \right) \end{equation} and
\begin{eqnarray} \label{c-and-d-coeff1} c(x_{m})&:=&\sqrt{\frac{
\pi x_{m}}{8}} \left[ \left(1+\frac{i}{2x_{m}}\right)H_{0}(x_{m})-
iH_{1}(x_{m})\right], \\ d(x_{m})&:=& \sqrt{\frac{\pi x_{m}}{8}}
\left[\left(1+\frac{i}{2x_{m}}\right)H^{*}_{0}(x_{m})
-iH^{*}_{1}(x_{m})\right].\label{c-and-d-coeff2}\end{eqnarray} Since
\begin{equation} |r_{0}(t)|^{2}-|s_{0}(t)|^{2}=1, \qquad
|c(x_{m})|^{2}-|d(x_{m})|^{2}=1,\end{equation} for all $t>0$ and
$m\in \mathbb{N}-\{0\}$, the maps $W_0(t)$ and $W(x_m)$ are
Bogoliubov transformations. Hence, the map from $S$ to $\Gamma$ is a
time dependent canonical transformation. A generating
functional for this transformation (that depends on some
appropriately chosen complete sets of compatible components -under
Poisson brackets- both for $S$ and $\Gamma$) is
$F=\sum_{m\in\mathbb{N}}F_{m}$ with
\begin{eqnarray} \label{generating-func0}
F_{0}(t)&=&
-\frac{i}{2r_{0}(t)}\left[s_{0}^{*}(t)b_{0}b_{0}-s_{0}(t)A_{0}^{*}
A_{0}^{*}+2b_{0}A_{0}^{*}\right],\\ \label{generating-func}
F_{m}(t)&=& ib_{-m}^{*}\left[c(x_{m})A_{-m}+
d(x_{m})A_{m}^{*}\right]- ib_{m}\left[
d^{*}(x_{m})A_{-m}+c^{*}(x_{m})A_{m}^{*}\right],\qquad m\neq 0.
\end{eqnarray}
A straightforward calculation shows that
$\partial_{t}F=\bar{H}_{r}$. Since the evolution in $S$ is frozen,
an initial state
$({\mathsf{B}}_{0}(t_{0}),\{{\cal{B}}_{m}(t_{0})\})$ in $\Gamma$ at
time $t_{0}$ will evolve to a state
$({\mathsf{B}}_{0}(t),\{{\cal{B}}_{m}(t)\})$ at time $t$ according
to
\begin{equation}  {\mathsf{B}}_{0}(t)  = W_{0}(t)W_{0}(t_{0})^{-1}
{\mathsf{B}}_{0}(t_{0}), \qquad \label{rel-dif-time}{\cal{B}}_{m}(t)
= W(x_{m})W(x_{m}^{0})^{-1}{\cal{B}}_{m}(t_{0}), \end{equation}
where $x_{m}^{0}:=mt_{0}$. In other words, the transformation
(\ref{rel-dif-time}) is the integral curve of the Hamiltonian vector
field $X^{A}=\Lambda^{AB}\nabla_{B}\bar{H}_{r}$ on $\Gamma$, with
end points at $({\mathsf{B}}_{0}(t_{0}),\{{\cal{B}}_{m}(t_{0})\})$
and $({\mathsf{B}}_{0}(t),\{{\cal{B}}_{m}(t)\})$. Alternatively, it
can also be viewed as the map that relates copies of $\Gamma$ at
different times, e.g.
$\big\{({\mathsf{B}}_{0}(t_{0}),\{{\cal{B}}_{m}(t_{0})\})\big\}$ at
$t_{0}$ with
$\big\{({\mathsf{B}}_{0}(t),\{{\cal{B}}_{m}(t)\})\big\}$ at $t$.

\section{Quantum time evolution}
\label{sec:5}
\setcounter{equation}{0}

Given a Cauchy surface in $({\cal{M}}\simeq \mathbb{R}^{+}\times
T^{2},g^{(S)})$, for instance the surface $t=t_{0}$, one obtains a
one-to-one correspondence between the spaces $S$ and $\Gamma$ by
means of the maps $W_{0}(t_{0})$ and $W(mt_{0})$ [see Eqs.
(\ref{zero-modes-map}) and (\ref{nonzer-modes-map})]. Employing the
inverse of these maps at $t=t_{0}$, one can then rewrite expression
(\ref{scalarf-in-a-var}) in terms of a new set of ``orthonormal"
mode solutions $\{G_{n}(t,\theta),G_{n}^*(t,\theta)\}$,
\begin{equation} \label{scalarf-ref-to}
\zeta(t,\theta) = \sum_{n=-\infty}^{\infty}
\left[G_{n}(t,\theta)b_{n}(t_{0})+G_{n}^{*}(t,\theta)
b_{n}^{*}(t_{0}) \right].
\end{equation}
For the zero and nonzero ($n\neq 0$) modes, respectively, these new
mode solutions are given by
\begin{eqnarray} \label{mode-sol-t00}
G_{0}(t,\theta)&=&\sqrt{t}\left[r_{0}^{*}(t_{0})f_{0}(t,\theta)-
s_{0}^{*}(t_{0})f^{*}_{0}(t,\theta)\right], \\
G_{n}(t,\theta)&=&\sqrt{\frac{t}{8}}
\left[c^{*}\big(x_{|n|}^{0}\big)
H_0(x_{|n|})-d^{*}\big(x_{|n|}^{0}\big) H^*_0(x_{|n|})\right]e^{i n
\theta}. \label{mode-sol-t0}
\end{eqnarray} Here, $x_{|n|}=|n| t$ and $x_{|n|}^0= |n|
t_0$.

Instead of $t=t_{0}$, we could have considered a different Cauchy
surface $t=T$. In such a case, the field $\zeta(t,\theta)$ would
have adopted an expression similar to (\ref{scalarf-ref-to}), but
now in terms of the set of coefficients $\{b_{n}(T),b_{n}^{*}(T)\}$
and the ``orthonormal" mode solutions,
$\{G^{(T)}_{n}(t,\theta),G^{(T)*}_{n}(t,\theta)\}$ that are obtained
by replacing $t_{0}$ with $T$ in Eqs.
(\ref{mode-sol-t00})-(\ref{mode-sol-t0}). Therefore, associated with
a uniparametric family (UF) of Cauchy surfaces $T\in
[t_{0},t_{f}]\subset \mathbb{R}^+$, there exists a UF of
``orthonormal" mode solutions
$\{G^{(T)}_{n}(t,\theta),G^{(T)*}_{n}(t,\theta)\}_{T\in
[t_{0},t_{f}]}$, as well as a UF of copies of $\Gamma$, namely
$\big\{({\mathsf{B}}_{0}(T),\{{\cal{B}}_{m}(T)\})\big\}_{T\in
[t_{0},t_{f}]}$, which are related via the evolution map
(\ref{rel-dif-time}).

Denoting $\bar{G}_{n}(t):=G_{n}(t,\theta)\exp[-i n \theta]$, the
explicit decomposition of the solutions in complex conjugate pairs
provided by Eq. (\ref{scalarf-ref-to}) allows one to introduce the
following $\Omega$-compatible complex structure:
\begin{equation} \label{our-cs} J\big[\bar{G}_{n}(t)\big] =
i\bar{G}_{n}(t) ,\qquad
J\big[\bar{G}^{*}_{n}(t)\big] = -i\bar{G}^{*}_{n}(t).
\end{equation} Given a UF of Cauchy surfaces,
we will also get a UF of $\Omega$-compatible complex structures
$J_{T}$, namely those defined by the UF  of ``orthonormal" mode
solutions $\{G^{(T)}_{n}(t,\theta),G^{(T)*}_{n}(t,\theta)\}_{T\in
[t_{0},t_{f}]}$:
\begin{equation} \label{ind-cs}
J_{T}\big[\bar{G}^{(T)}_{n}(t)\big]=i\bar{G}^{(T)}_{n}(t), \qquad
J_{T}\big[\bar{G}^{(T)*}_{n}(t)\big]=-i\bar{G}^{(T)*}_{n}(t),
\end{equation} where again
$\bar{G}^{(T)}_{n}(t):=G^{(T)}_{n}(t,\theta)\exp[-i n \theta]$.
Thus, for each copy
$\big\{({\mathsf{B}}_{0}(T),\{{\cal{B}}_{m}(T)\})\big\}$ of
$\Gamma$, we obtain a natural complex structure $J_{T}:S \to S$.
Since the copies of $\Gamma$ are related by the evolution map
(\ref{rel-dif-time}), the UF $\{J_{T}\}_{T\in [t_{0},t_{f}]}$ is
just the set of complex structures induced by time evolution.
Obviously, $J=J_{T}|_{T=t_{0}}$ and
$G_{n}(t,\theta)=G^{(T)}_{n}(t,\theta)|_{T=t_{0}}$.

Starting with $(S,J)$, we can construct the ``one-particle" Hilbert
space ${\cal{H}}$. It is the (Cauchy) completion of the space of
``positive frequency" solutions
$S^{+}:=\{\zeta^{+}=(\zeta-iJ\zeta)/2\}$ with respect to the norm
$||\zeta^{+}||=\sqrt{\langle\zeta^{+} , \zeta^{+} \rangle}$. Here,
$\langle \cdot , \cdot \rangle$ denotes the Klein-Gordon inner
product: $\langle \zeta^{+} , \varsigma^{+} \rangle =
-i\Omega(\zeta^{-},\varsigma^{+})$ where $\zeta^{-}\in
\bar{{\cal{H}}}$ (i.e., the complex conjugate space of ${\cal{H}}$).
The (kinematical) Hilbert space of the quantum theory is the
symmetric Fock space ${\cal{F}}({\cal{H}})$ constructed from the
``one-particle" Hilbert space. That is,
\begin{equation} {\cal{F}}({\cal{H}})=\oplus _{k=0}^{\infty}
\left(\otimes^{k}_{(s)} {\cal{H}}\right) , \end{equation} where
$\otimes^{k}_{(s)} {\cal{H}}$ is the Hilbert space of all $k$th rank
symmetric tensors over $\cal{H}$. Following this prescription, we
can write the formal field operator $\hat{\zeta}$ in terms of
annihilation and creation operators corresponding to the positive
and negative frequency decomposition defined by the complex
structure $J$: \begin{equation} \label{field-op}
\hat{\zeta}(t;\theta)=\sum_{n=-\infty}^{\infty}
\left[G_{n}(t,\theta)\hat{b}_{n}+
G_{n}^{*}(t,\theta)\hat{b}_{n}^{\dagger}\right]. \end{equation}
Remembering expression (\ref{scalarf-ref-to}) for the classical
solutions, we see that we might have obtained this field operator by
a straightforward assignation of operators to constants of motion.
This is the Schr\"{o}dinger picture, where the complex constant
coefficients $\{b_{n}(t_{0}),b_{n}^{*}(t_{0})\}$ are promoted to
annihilation and creation operators
$\{\hat{b}_{n}(t_{0})=\hat{b}_{n},\hat{b}^{\dagger}_{n}(t_{0})=
\hat{b}^{\dagger}_{n}\}$.

We note that $\{(S,J_{T})\}_{T\in \mathbb{R}^{+}}$ leads to the UF
of Fock representations
$\left\{({\cal{F}}({\cal{H}}_{T}),\{\hat{b}_{n}(T),
\hat{b}_{n}^{\dagger}(T)\})\right\}_{T\in \mathbb{R}^{+}}$. Clearly,
the Fock representation constructed from $(S,J)$ belongs to this
family and corresponds to $T=t_{0}$.

On the other hand, in the Heisenberg picture, time evolution for
operators is determined by the Bogoliubov transformation
(\ref{rel-dif-time}). Introducing then the notation
$\hat{b}^{(H)}_{n}(t_{0}):=\hat{b}_{n}$, one obtains the following
relation between the annihilation and creation operators at $t_0$
and a different time $t$:
\begin{equation} \label{b-quant-evolution}
\hat{b}^{(H)}_{n}(t)=\alpha_{n}(t,t_{0})\hat{b}^{(H)}_{n}(t_{0})
+\beta_{n}(t,t_{0})\hat{b}_{-n}^{(H)\dagger}(t_{0}),
\end{equation} where the Bogoliubov coefficients for the zero modes
are [see Eq. (\ref{c-and-d-coeff0})]\begin{equation}
\alpha_{0}(t,t_{0})=r_{0}(t)r^{*}_{0}(t_{0})-
s_{0}(t)s^{*}_{0}(t_{0}), \qquad
\beta_{0}(t,t_{0})=s_{0}(t)r_{0}(t_{0})-
r_{0}(t)s_{0}(t_{0}),
\end{equation}
while for the rest of modes with $n\in \mathbb{Z}-\{0\}$ one gets
[see Eqs. (\ref{c-and-d-coeff1}) and (\ref{c-and-d-coeff2})]
\begin{equation}\label{beta}
\alpha_{n}(t,t_{0})=c(x_{|n|})c^{*}(x^{0}_{|n|}) -d(x_{|n|})
d^{*}(x^{0}_{|n|}), \qquad \beta_{n}(t,t_{0})=d(x_{|n|})
c(x^{0}_{|n|})-c(x_{|n|})d(x^{0}_{|n|}).
\end{equation}

Interchanging the roles of $t$ and $t_{0}$ in Eq.
(\ref{b-quant-evolution}), we can write the annihilation operator at
time $t_{0}$ in terms of the annihilation and creation operators at
time $t$. By substituting the result in Eq.
(\ref{b-quant-evolution}), and using the relations
$|\alpha|^{2}-|\beta|^{2}=1$,
$\alpha_{n}(t,t_{0})=\alpha_{n}^{*}(t_{0},t)$, and
$\beta_{n}(t,t_{0})=-\beta_{n}(t_{0},t)$, we then get that the
creation operator at time $t_{0}$ is equal to
$\beta^{*}_{n}(t_{0},t)\hat{b}^{(H)}_{-n}(t)
+\alpha^{*}_{n}(t_{0},t)\hat{b}_{n}^{(H)\dagger}(t)$. Hence, the
field operator can be fully expressed in terms of operators in the
Heisenberg picture:
\begin{equation}
\hat{\zeta}(t;\theta)=\frac{1}{\sqrt{4\pi}} \left(
\hat{b}^{(H)}_{0}(t)+\hat{b}_{0}^{(H)\dagger}(t)\right)
+\sum_{n=-\infty, n\neq 0}^{\infty} \frac{1}{\sqrt{4\pi|n|}} \left(
e^{in\theta}\hat{b}^{(H)}_{n}(t)+e^{-in\theta}
\hat{b}_{n}^{(H)\dagger}(t)\right).
\end{equation}

\section{Unitarity of the evolution}
\label{sec:6}
\setcounter{equation}{0}

The time evolution in the Heisenberg picture described in the
previous section is unitarily implementable on the (kinematical)
Fock space ${\cal{F}}({\cal{H}})$ constructed from $(S,J)$ if and
only if the sequence $\{\beta_{n}(t,t_{0})\}$ that appears in
relation (\ref{b-quant-evolution}) is square summable \cite{shale}
(see also Ref. \cite{honegger}). Let us remark that unitary
implementability amounts to unitary equivalence between all the Fock
representations in the UF under consideration. Since two complex
structures $J_{T_{1}}$ and $J_{T_{2}}$ lead to unitary equivalent
representations of the canonical commutation relations if and only
if their difference $(J_{T_{1}}-J_{T_{2}})$ defines a
Hilbert-Schmidt (HS) operator, either on ${\cal{H}}_{T_{1}}$ or
${\cal{H}}_{T_{2}}$ (see e.g. \cite{ash-ma}), we have unitary
implementability if and only if ${\cal{J}}_{T}:=(J-J_{T})$ is HS
for every $T\in \mathbb{R}^{+}$. In fact, it is not difficult to see
that ${\cal{J}}_{T}$ is HS if and only if the sequence
$\{\beta_{n}(T,t_{0})\}$ is square summable (as it should be,
because the requirement of unitary equivalence between Fock
representations is just a reformulation of the unitary
implementability condition).

Let us discuss then the square summability of the sequence
$\{\beta_{n}(t,t_{0})\}$. Since
$\beta_{n}(t,t_{0})=\beta_{-n}(t,t_{0})$ and, in addition,
summability does not depend on the contribution of a single term
(e.g. $n=0$), it suffices to analyze the sequence
$\{\beta_{m}(t,t_{0})\}$ with $m\in \mathbb{N}-\{0\}$. We start by
showing that the sequence $\{d(mt)\}$ $(m\in \mathbb{N}-\{0\})$ is
square summable for all $t>0$. From the asymptotic expansions of
the Hankel functions for large positive arguments, we know that
\cite{abramowitz}
\begin{eqnarray} H_{0}(x)&=&\sqrt{\frac{2}{\pi x}}\left[P(0,x)-
iQ(0,x)\right]e^{-ix} e^{i\pi/4}, \label{asympt-hankel0} \\
H_{1}(x)&=&\sqrt{\frac{2}{\pi x}}\left[Q(1,x)+ iP(1,x)\right]
e^{-ix} e^{i\pi/4}, \label{asympt-hankel1}
\end{eqnarray}
where
\begin{eqnarray} P(\nu,x)&:=&
1+p(\nu,x)\sim 1+\sum_{k=1}^{\infty}(-1)^{k}\frac{(\nu,2k)}
{(2x)^{2k}},
\\ Q(\nu , x)&\sim & \sum_{k=0}^{\infty}(-1)^{k}\frac{(\nu,2k+1)}
{(2x)^{2k+1}}.
\end{eqnarray} Here, $\nu=0,1$ and $(\nu,m)$ are Hankel's
symbols, which in terms of the gamma function are
\begin{equation}\label{hansym}
(\nu,k)=\frac{\Gamma\big(\nu+k+\frac{1}{2}\big)}
{k!\Gamma\big(\nu-k+\frac{1}{2}\big)}
.\end{equation} Substituting Hankel's expansions in Eq.
(\ref{c-and-d-coeff2}), the square modulus of $d(x_{m})$ in the
asymptotic region $x_{m}\gg 1$ becomes
\begin{equation}
|d(x_{m})|^{2}=\left(\frac{p(0,x_{m})-p(1,x_{m})}{2}
-\frac{Q(0,x_{m})}
{4x_{m}}\right)^{2}+\left(\frac{Q(0,x_{m})-Q(1,x_{m})}{2}+
\frac{1+p(0,x_{m})}{4x_{m}}\right)^{2}.
\end{equation} Employing the expressions of the functions
$p(\nu,x)$ and $Q(\nu,x)$, it is a simple exercise to check that the
last term in round brackets is $o(1/x_m^5)$ at infinity\footnote{We
say that a function $f(x)$ is $o(1/x^n)$ when $x\rightarrow \infty$
if the product $x^n f(x)$ tends to zero in this limit.}, whereas the
first one presents the behavior $1/(4x_m)^4+o(1/x_m^5)$. So, for
$x_m\gg 1$ one gets
\begin{equation}\label{dsquare} |d(x_{m})|^{2}=\frac{1}{(4
x_m)^4}+o\left(\frac{1}{x_m^5}\right).\end{equation} This implies,
for instance, that one can find a positive constant $C$ so that
$|d(x_{m})|^{2}\leq 1/(3 x_m)^4$ if $x_m>C$ [because  $1/(3)^4>
1/(4)^4$]. For every given value of $t>0$, let us call $M_0:={\rm
int}{(C/t)}<\infty$, where ${\rm int}{(x)}$ denotes the integer part
of $x$. We then obtain
\begin{equation} \sum_{m=1}^{\infty}|d(mt)|^{2}\leq
\sum_{m=1}^{M_{0}}|d(mt)|^{2}+\frac{1}{(3t)^{4}}
\sum_{m=M_{0}+1}^{\infty}\frac{1}{m^{4}}<\infty.
\end{equation} In the last inequality we have used that the
first sum involves only a finite number of (well defined and
bounded) terms and that the sequence $\{1/m^4\}$ is summable.
This proves that the sequence $\{d(mt)\}$ with $m\in
\mathbb{N}-\{0\}$ is square summable for all positive times $t>0$.

Given two values of the time coordinate, $t>0$ and $t_0>0$, we know
from the square summability of the sequences $\{d(mt)\}$ and
$\{d(mt_{0})\}$ [and without appealing to the explicit form of
$d(x)$] that there exist two integers $m_{0}$ and $\tilde{m}_{0}$
such that $|d(mt)|<1$ for all $m>m_{0}$ and $|d(mt_{0})|<1$ for all
$m>\tilde{m}_{0}$. Remembering that $|c|^{2}=1+|d|^{2}$, one also
has that $|c(mt)|$ and $|c(mt_{0})|$ are smaller than $\sqrt{2}$ for
all $m>M_{1}:=\max (m_{0},\tilde{m}_{0})$. In this case, we obtain
from Eq. (\ref{beta})
\begin{equation} |\beta_{m}(t,t_0)| \leq
\sqrt{2}\left(|d(x_{m})|+|d(x^{0}_{m})|\right). \end{equation} Using
the inequality
\begin{equation} \left(|d(x_{m})|+|d(x^{0}_{m})|\right)^{2}\leq
2\left(|d(x_{m})|^{2}+|d(x^{0}_{m})|^{2}\right), \end{equation} we
then conclude that
\begin{equation} \label{bound-seq} \sum_{m=1}^{\infty}|
\beta_{m}(t,t_0)|^{2}\leq \sum_{m=1}^{M_{1}}|\beta_{m}(t,t_0)|^{2} +
4\sum_{m=M_{1}+1}^{\infty}\left(|d(x_{m})|^{2}+
|d(x^{0}_{m})|^{2}\right). \end{equation} Provided that
$|\beta_{m}(t,t_0)|$ is bounded from above for all $t,t_{0}>0$ and
$m\in \mathbb{N}-\{0\}$, the first sum in the right hand side
of Eq. (\ref{bound-seq}) is bounded since it contains only a finite
number of contributions. The second sum is bounded from above as
well, because both sequences $\{d(x_{m})\}$ and $\{d(x^{0}_{m})\}$
are square summable. Therefore, the sequence
$\{\beta_{m}(t,t_{0})\}$ is square summable for all $t,t_{0}>0$.

As an aside, let us comment that Hankel's expansions
(\ref{asympt-hankel0}) and (\ref{asympt-hankel1}) provide the
following asymptotic behavior of the Bogoliubov coefficient
$\beta_{m}(t,t_{0})$ for large $m$:
\begin{equation} \beta_{m}(t,t_0)=\frac{1}{16m^{2}}\left[
\left(\frac{1} {(t-\Delta t)^{2}}-\frac{1}{t^{2}}\right)\cos(m\Delta
t)-i \left(\frac{1}{(t-\Delta t)^{2}}+\frac{1}{t^{2}}\right)
\sin(m\Delta t)\right] + o\left(\frac{1}{m^{2}}\right),
\end{equation} where $\Delta t := t-t_{0}$. Obviously, in agreement
with our above comments and in contrast to the situation found in
Eq. (\ref{approx-anti-old}), the dominant term in the asymptotic
regime is square summable. Moreover, it is not difficult to see
that, when $t>> 1$ and $\Delta t /t << 1$, the above expression
continues to be valid for all $m>0$ if one merely replaces
$o(1/m^2)$ with $o(1/t^2)$. Approximating to the same order the rest
of terms in the expression, we arrive at a square modulus for
$\beta_{m}(t,t_0)$ with the asymptotic form
\begin{equation}
|\beta_{m}(t,t_0)|^{2}=\frac{\sin^{2}(m\Delta t)}{64(mt)^{4}}+
o\left(\frac{1}{t^4}\right).
\end{equation}
For asymptotic large times
we see that, if we keep $\Delta t$ fixed, the square modulus of
$\beta_{m}(t,t_0)$ decreases to zero as $1/t^4$. Since the number of
``particles'' produced by the vacuum in the nonzero modes is
given by the sum of the sequence
$\big\{|\beta_{m}(t,t_0)|^{2}\big\}$, in a fixed lapse of time
$\Delta t$ the ``particle'' production will be attenuated as time
increases. In particular, regardless of the fixed value of $\Delta
t$, one gets the bound
\begin{equation} \sum_{m=1}^{\infty}|\beta_{m}(t,t_0)|^{2}\leq
\sum_{m=1}^{\infty}
\frac{1}{64m^{4}t^4}+o\left(\frac{1}{t^4}\right)=\frac{1}
{64t^4}Z(4)+o\left(\frac{1}{t^4}\right),\end{equation} where
$Z(4)=\pi^{4}/90$ is the Riemann function $Z(j)$ at $j=4$. We will
discuss the issue of ``particle'' production in more detail in the
next section.

The proved square summability of the sequence $\{\beta_{m}(t,t_0)\}$
for all positive times $t$ and $t_0$ ensures that the time evolution
is unitarily implementable on the (kinematical) Fock space
${\cal{F}}({\cal{H}})$ [and that the Fock representations of the
introduced UF are all unitarily equivalent], so that probability is
preserved. Moreover, one can check that the evolution
(\ref{b-quant-evolution}) leaves invariant the constraint
\begin{equation} \hat{C}_{0}=\sum_{m=1}^{\infty} m
\left(\hat{b}^{\dagger}_{m}\hat{b}_{m}-
\hat{b}^{\dagger}_{-m}\hat{b}_{-m}\right), \end{equation} which
imposes the condition that the total momentum of the field $\xi$ in
the $\theta$-direction be equal to zero [see Eq.
(\ref{gconst-b-var})]. This invariance guarantees that the dynamics
is unitarily implementable not just on ${\cal{F}}({\cal{H}})$, but
furthermore on the Hilbert space ${\cal{F}}_{\rm{phys}}({\cal{H}})$
of physical states, which are the states that belong to the kernel
of the constraint $\hat{C}_0$. In conclusion, we have shown that the
quantization put forward for the polarized Gowdy $T^3$ model is such
that the physical evolution is unitary.

\section{Features and consequences of the quantum evolution}
\label{sec:7}
\setcounter{equation}{0}

In this section, we want to clarify certain mathematical aspects of
the quantization and evolution proposed for the Gowdy model and
discuss some of their physical consequences, including the
cosmological production of ``particles'' by the vacuum of the
theory. We divide this analysis in several parts.

(1) Associated with the field decomposition (\ref{scalarf-in-a-var})
we have the $\Omega$-compatible complex structure
\begin{equation} J_g\big[\bar{g}_{n}(t)\big]=i\bar{g}_{n}(t), \qquad
J_g\big[\bar{g}^{*}_{n}(t)\big]=-i\bar{g}^{*}_{n}(t),
\end{equation}
where $\bar{g}_{n}(t):=g_{n}(t,\theta)\exp[-in\theta]$. Starting
with $(\Omega,\{\zeta\},J_g)$ but adopting
$({\mathsf{A}}_{0},\{{\cal{A}}_{m}\})$ as coordinates for
$\{\zeta\}$ instead of
$({\mathsf{B}}_{0}(t_{0}),\{{\cal{B}}_{m}(t_{0})\})$, one can
construct the ``one-particle" Hilbert space ${\cal{H}}_g$ as well as
the corresponding symmetric Fock space ${\cal{F}}({\cal{H}}_g)$.
This Fock space would now provide the (kinematical) Hilbert space of
the quantum theory. Defining ${\cal{J}}:=J-J_g$, we have that
\begin{equation} \frac{1}{4}\sum_{n=-\infty}^{\infty}\langle
G_{n},{\cal{J}}^{\dagger}{\cal{J}}[
G_{n}]\rangle_{{\cal{H}}}=|s_{0}(t_{0})|^{2}+2\sum_{m=1}^{\infty}
|d(x^{0}_{m})|^{2}<\infty. \end{equation} We can therefore assure
that
$({\cal{F}}({\cal{H}}_g),\{\hat{A}_{n},\hat{A}^{\dagger}_{n}\})$ and
$({\cal{F}}({\cal{H}}),\{\hat{b}_{n},\hat{b}^{\dagger}_{n}\})$ are
unitarily equivalent Fock representations. Besides, since the
sequence $\{d(x_{m})\}$ is square summable for all $t>0$, the
unitary equivalence holds regardless of the value chosen for the
instant $t_{0}>0$ in the construction of the representation
$({\cal{F}}({\cal{H}}),\{\hat{b}_{n}=\hat{b}_{n}(t_{0}),
\hat{b}^{\dagger}_{n}=\hat{b}^{\dagger}_{n}(t_{0})\})$. In this
sense, the role of the time of reference $t_0$ is irrelevant.

In the considered description, on the other hand, an initial state
$({\mathsf{A}}_{0},\{{\cal{A}}_{m}\})$ at time $t_{0}$ evolves to
the final state $(\bar{{\mathsf{A}}}_{0},\{\bar{{\cal{A}}}_{m}\})$
at time $t$ according to
\begin{equation} \label{a-evol}
\bar{{\mathsf{A}}}_{0}= W_{0}^{-1}(t_{0})
W_{0}(t){\mathsf{A}}_{0}, \qquad
\bar{{\cal{A}}}_{m} = W^{-1}(x^{0}_{m})W(x_{m}){\cal{A}}_{m}.
\end{equation}
The antilinear part of the map (\ref{a-evol}) is obviously square
summable and, consequently, the classical dynamics is unitarily
implementable with respect to the Fock representation
$({\cal{F}}({\cal{H}}_{g}),\{\hat{A}_{n},\hat{A}^{\dagger}_{n}\})$,
as required for consistency with the unitary equivalence between
this representation and
$({\cal{F}}({\cal{H}}),\{\hat{b}_{n},\hat{b}^{\dagger}_{n}\})$.

However, it is worth pointing out that the transformation
(\ref{a-evol}) does not represent the total change in time, which is
actually dictated by $K:={\bar{H}}_{r}-\partial_{t}F=0$ [see Eqs.
(\ref{generating-func0}) and (\ref{generating-func})]. It is rather
the relation between constants of motion that generates the
Hamiltonian (\ref{hamil-in-b-coord}), written in coordinates
$({\mathsf{A}}_{0},\{{\cal{A}}_{m}\})$. This situation contrasts
with that described in Eq. (\ref{rel-dif-time}), where the total
Hamiltonian is indeed ${\bar{H}}_{r}$. This observation is one of
the main motivations for the construction of the Fock representation
$({\cal{F}}({\cal{H}}),\{\hat{b}_{n},\hat{b}^{\dagger}_{n}\})$ that
we have presented, representation where the total dynamics provided
by Eq. (\ref{rel-dif-time}) is implemented in a natural way.

(2) Let us analyze now the regime of times $T\in
[\tilde{T},\infty)$, with $\tilde{T}$ large enough so that
$d(|n|\tilde{T})$ can be neglected with respect to the unity in an
asymptotic approximation. We will then have $d(|n|T)\approx 0$ and
$c(|n|T)\approx \exp[i\delta_{|n|}(T)]$ where, for each nonzero
integer $|n|$, the phase $\delta_{|n|}(T)$ is some smooth real
function of $T$. Therefore, for the nonzero modes, the solutions
$G^{(T)}_{n}(t,\theta)$ considered in Sec. V behave at leading order
as
\begin{equation} G^{(T)}_{n}(t,\theta)\approx
\sqrt{\frac{t}{8}}e^{in\theta}e^{-i\delta_{|n|}(T)}H_{0}(|n|t)
\end{equation} for $T\geq \tilde{T}$. From Hankel's asymptotic
expansion (\ref{asympt-hankel0}) of $H_{0}$, we also get that, for
large values of $t$,
\begin{eqnarray} \label{asympt-sol} G^{(T)}_{n}(t,\theta) &
\approx & \eta_{|n|}(T)e^{-i(|n|t-n\theta)}, \\
\eta_{|n|}(T) & := &
\frac{1}{2\sqrt{\pi|n|}}e^{-i\delta_{|n|}(T)}e^{i\pi/4}.
\end{eqnarray}

Hence, in the asymptotic region of large times, solutions
(\ref{asympt-sol}) approach the (nonzero) ``orthonormal" mode
solutions for a free massless scalar field propagating in the static
background $({\cal{M}}\simeq \mathbb{R}^{+}\times T^{2},g^{(S)})$
[see Eq. (\ref{stabac})]. From Eqs. (\ref{ind-cs}) and
(\ref{asympt-sol}) it follows that, in the limit in which the system
becomes massless, $J_{T}$ approaches the (counterpart of the)
Poincar\'{e}-invariant complex structure of Minkowski spacetime,
namely $J_{M}=-(-\pounds_{t}\pounds_{t})^{-1/2}\pounds_{t}$, where
$\pounds_{t}$ is the Lie derivative along $t^{a}:= (\partial /
\partial t)^{a}$. In fact, this is not unexpected: for
asymptotically large values of $t$, the Hamiltonian
(\ref{hamil-in-b-coord}) describes (in the sector of nonzero modes)
a collection of harmonic oscillators with frequency
$\omega_{n}=|n|$, as we have already seen.

(3) In Secs. V and VI, we have formulated the quantum time evolution
in the Heisenberg picture. For completeness, we will now discuss the
Schr\"{o}dinger picture. In this picture, the evolution is attained
by implementing the Bogoliubov transformation (\ref{rel-dif-time})
on the ``one-particle" Hilbert space ${\cal{H}}$
\cite{torre-vara-fevol}. Namely, the initial state
$\zeta^{+}(t_{0})=\sum_{n\in \mathbb{Z}}G_{n}(t,\theta)b_{n}
(t_{0})$ at $t=t_{0}$ will evolve to the final state
$\zeta^{+}(t_f)=\sum_{n\in \mathbb{Z}}G_{n}(t,\theta)b_{n}(t_{f})$
at $t=t_{f}$, with $b_{n}(t_{0})$ and $b_{n}(t_{f})$ related by Eq.
(\ref{rel-dif-time}). This transformation defines a pair of bounded
linear maps $\alpha : {\cal{H}}\to {\cal{H}}$ and $\beta
:{\cal{H}}\to \bar{{\cal{H}}}$ (recall that $\bar{{\cal{H}}}$ is the
complex conjugate space),
\begin{eqnarray} \alpha \cdot \zeta^{+}(t_{0}) & = &
\sum_{n=-\infty}^{\infty}G_{n}(t,\theta)
\alpha_{n}(t,t_{0})b_{n}(t_{0}),  \\
\beta \cdot \zeta^{+}(t_{0}) & = &
\sum_{n=-\infty}^{\infty}G^{*}_{-n}(t,\theta)
\beta^{*}_{n}(t,t_{0})b_{n}(t_{0}).
\end{eqnarray} Unitary implementability is possible if and only if the
operator $\beta$ is Hilbert-Schmidt, that is, if and only if
\begin{eqnarray} {\rm{tr}} \left(\beta^{\dagger}\beta\right) &= &
\sum_{n=-\infty}^{\infty} \langle G_{n}, \beta^{\dagger}\beta \cdot
G_{n} \rangle_{{\cal{H}}} = \sum_{n=-\infty}^{\infty}\langle \beta
\cdot G_{n}, \beta \cdot G_{n}
\rangle_{\bar{\cal{H}}} \nonumber \\
& = & \sum_{n=-\infty}^{\infty}\langle \beta_{n}^{*} G^{*}_{-n},
\beta_{n}^{*} G^{*}_{-n} \rangle_{\bar{\cal{H}}} =
\sum_{n=-\infty}^{\infty} |\beta_{n}|^{2} < \infty.  \end{eqnarray}
As we have shown, this is in fact the case.

It should be stressed that the maps $\alpha$ and $\beta$ define
the unitary map ${\cal{U}}$ that implements the dynamics on
${\cal{F}(\cal{H})}$. Namely, considering the standard annihilation
operator associated to $\zeta^{+}(t_{0})$ \cite{wald}, i.e. the
smeared annihilation operator
$\hat{\mathfrak{b}}\big(\overline{\zeta^{+}(t_{0})}\big)$, we know
that ${\cal{U}}$ is defined -up to a phase- by \cite{honegger,wald}
\begin{equation}
\label{smea-ev}
{\cal{U}}(t_{f},t_{0})\hat{\mathfrak{b}}
\big(\overline{\zeta^{+}(t_{0})}\big){\cal{U}}^{\dagger}
(t_{f},t_{0})= \hat{\mathfrak{b}}
\big(\overline{\alpha\cdot \zeta^{+}(t_{0})}\big)-
\hat{\mathfrak{b}}^{\dagger}\big(\overline{\beta\cdot
\zeta^{+}(t_{0})}\big).
\end{equation}
Since the antilinear part $\beta$ of the Bogoliubov transformation
is not null, the vacuum [the state $|0,t_{0}) \in
{\cal{F}(\cal{H})}$ annihilated by $\hat{\mathfrak{b}}(\bar{\eta})$
for all $\eta \in {\cal{H}}$] does not remain invariant under the
action of ${\cal{U}}$. That is, ${\cal{U}}(t_{f},t_{0})|0,t_{0})$
will not be annihilated by $\hat{\mathfrak{b}}(\bar{\eta})$ for all
$\eta \in {\cal{H}}$. Note that Eq. (\ref{smea-ev}) is just the
smeared version of Eq.(\ref{b-quant-evolution}). In terms of the
considered UF of Fock representations, the operator (\ref{smea-ev})
can be viewed as the annihilation operator in the $t_{f}$ Fock space
(that associated with the Cauchy surface $t_{f}$). Of course, the
vacuum state of the $t_{f}$ Fock representation will not coincide
with $|0,t_{0})$, but rather be given by
$|0,t_{f})={\cal{U}}(t_{f},t_{0})|0,t_{0})$.

Some additional comments may be worth at this stage. As we
have said, the state ${\cal{U}}(t_{f},t_{0})|0,t_{0}) \in
{\cal{F}(\cal{H})}$ can be identified as the vacuum of the $t_{f}$
Fock space. Thus, the evolution map ${\cal{U}}(t_{f},t_{0})$ in
${\cal{F}(\cal{H})}$ can also be viewed as the unitary map relating
the $t_{0}$ and $t_{f}$ Fock representations. On the other hand, in
order to determine the evolution in the dense subspace of states
with a finite number of particles, one only needs to know how the
``$n-$particle" states evolve, and this in turn becomes completely
fixed by specifying how the vacuum and the creation operators change
in time: given a ``$n-$particle" state
$|n_{\zeta})=\hat{\mathfrak{b}}^{\dagger}(\zeta_{1}^{+})
\hat{\mathfrak{b}}^{\dagger}(\zeta_{2}^{+}) \dots
\hat{\mathfrak{b}}^{\dagger}(\zeta_{n}^{+})|0,t_{0})$ [and
abbreviating ${\cal{U}}(t_{f},t_{0})$ to ${\cal{U}}$], one has
\begin{equation}
{\cal{U}}|n_{\zeta})={\cal{U}}\hat{\mathfrak{b}}^{\dagger}
(\zeta_{1}^{+}){\cal{U}}^{\dagger}{\cal{U}}
\hat{\mathfrak{b}}^{\dagger} (\zeta_{2}^{+})
{\cal{U}}^{\dagger}{\cal{U}}\dots
{\cal{U}}\hat{\mathfrak{b}}^{\dagger}(\zeta_{n}^{+})
{\cal{U}}^{\dagger}{\cal{U}}|0,t_{0}).
\end{equation}
Therefore, the adjoint of  Eq. (\ref{smea-ev}) and the corresponding
relation between vacua that provides
${\cal{U}}(t_{f},t_{0})|0,t_{0})$ determines indeed the evolution in
${\cal{F}(\cal{H})}$.

Similar comments apply to the $t_{0}$ Fock representation
constructed without smearing operators, i.e., the Fock
representation
$({\cal{F}}({\cal{H}}),\{\hat{b}_{n}(t_{0}),\hat{b}^{\dagger}_{n}
(t_{0})\})$. One only has to replace the smeared operators
$\hat{\mathfrak{b}}$ with the $\hat{b}_{n}$'s and notice that the
unitary operator is now defined by Eq.(\ref{b-quant-evolution}). Let
us explain this point in more detail, for the sake of clarity. In
the Fock representation
$({\cal{F}}({\cal{H}}),\{\hat{b}_{n}(t_{0}),\hat{b}^{\dagger}_{n}
(t_{0})\})$, the evolution operator is defined by
\begin{equation}
\label{unsmea-ev}
U(t_{f},t_{0})\hat{b}_{n}(t_{0})U^{\dagger}(t_{f},t_{0})=
\alpha_{n}(t_f,t_{0})\hat{b}_{n}(t_{0})
+\beta_{n}(t_f,t_{0})\hat{b}_{-n}^{\dagger}(t_{0}).
\end{equation}
The corresponding vacuum state $|0,t_{0}\rangle$
(characterized by the conditions
$\hat{b}_{n}(t_{0})|0,t_{0}\rangle = 0$ for all $n$) evolves to
the state $|0,t_{f}\rangle =
U(t_{f},t_{0})|0,t_{0}\rangle$, which corresponds in turn to the
vacuum of the $t_{f}$ Fock representation
$({\cal{F}}({\cal{H}}_{t_{f}}),\{\hat{b}_{n}(t_{f}),
\hat{b}^{\dagger}_{n}(t_{f})\})$, where the annihilation
operator $\hat{b}_{n}(t_{f})$ is given
 by Eq. (\ref{unsmea-ev}). Since
$\beta_{n}(t_{f},t_{0})$ does not vanish for $t_{f}\neq t_{0}$,
$|0,t_{0}\rangle$ and $|0,t_{f}\rangle$ do not simply differ by a
phase. The explicit relation between these vacua will be presented
below.

(4) Let us analyze now the issue of ``particle'' production,
focusing our discussion on the nonzero modes. Among the UF of Fock
representations $\left\{({\cal{F}}({\cal{H}}_{T}),\{\hat{b}_{n}(T),
\hat{b}^{\dagger}_{n}(T)\})\right\}_{T\in \mathbb{R}^{+}}$, let us
consider the $T=t_{0}$ and $T=t_{f}$ representations. The
expectation value of the number operator at time $t_{f}$, namely
$\hat{N}(t_{f})=\sum_{n\neq
0}\hat{b}^{\dagger}_{n}(t_{f})\hat{b}_{n}(t_{f})$, in the vacuum
state at time $t_{0}$, $|0,t_{0}\rangle$, is given by
\begin{equation} \langle 0,t_{0}
|\hat{N}(t_{f})|0,t_{0}\rangle=\sum_{n=-\infty, n\neq
0}^{\infty}|\beta_{n}(t_{f},t_{0})|^{2}=2\sum_{m=1}^{\infty}
|\beta_{m}(t_{f},t_{0})|^{2}.
\end{equation} This expectation value is different from zero, but
also bounded from above, because the sequence
$\{\beta_{m}(t_{f},t_{0})\}$ is square summable for all times
$t_{0},t_{f}>0$. As we have seen, for asymptotically large
values of $t_{0}$ and $t_{f}$ we can neglect the value of
$d(|n|t_{0})$ and $d(|n|t_{f})$, so that $\beta_{n}(t_{f},t_{0})
\approx 0$ and $\alpha_{n}(t_{f},t_{0})\approx
\exp(i[\delta_{|n|}(t_{f})-\delta_{|n|}(t_{0})])$. Then, in the
asymptotic region, $\hat{b}_{n}(t_{f})$ and $\hat{b}_{n}(t_{0})$
differ only by the phase
$\exp(i[\delta_{|n|}(t_{f})-\delta_{|n|}(t_{0})])$ [i.e.,
$(J_{t_{f}}-J)\approx 0$], and $\hat{N}(t_{0})\approx
\hat{N}(t_{f})$. That is, the ``particle'' production decreases as
$t_{0}$ and $t_{f}$ grow and, consequently, $|0,t_{f}\rangle \approx
|0,t_{0}\rangle$.

Actually, the evolution of the vacuum can be straightforwardly
calculated by remembering that the vacuum at time $T$ is
characterized (up to a phase) as the unit state annihilated by all
of the operators $\hat{b}_n(T)$. From the evolution of these
operators, it is then not difficult to see that the relation between
the studied vacua is
\begin{equation} \label{vac-rel}|0,t_{f}\rangle = F
\exp{\left[-\sum_{m=1}^{\infty}\lambda_{m}(t_{f},
t_{0})\hat{b}^{\dagger}_{m}(t_{0})\hat{b}^{\dagger}_{-m}
(t_{0})\right]} |0,t_{0}\rangle , \end{equation} where
$\lambda_{m}(t_{f},t_{0}):=\beta_{m}(t_{f},t_{0})/
\alpha_{m}(t_{f},t_{0})$ is the ratio between Bogoliubov
coefficients and $F$ is a normalization factor. Demanding that the
vacua have unit norm, we obtain
\begin{equation} \label{trans-prob} |F|
= \prod_{m=1}^{\infty}\sqrt{1-|\lambda_{m}
(t_{f},t_{0})|^{2}}.\end{equation} To the best of our knowledge
expressions (of the form of) (\ref{vac-rel}) and
(\ref{trans-prob}) were first derived, in a cosmological context,
almost forty years ago \cite{parker-2}. Notice that
$|F|^{2}=|\langle 0,t_{0}|0,t_{f}\rangle|^{2}$. That is, $|F|^{2}$
is the probability of finding no ``particles'' [corresponding to
$\hat{b}_{m}(t_{f})$] at time $t_{f}$, provided that the state
contains no ``particles'' at time $t_{0}$. Therefore,
$|\lambda_{m}(t_{f},t_{0})|^{2}$ gives the probability of observing
a nonzero number of ``particles'' in the mode $m$ (or $-m$) at
time $t_{f}$. For asymptotically large values of $t_{0}$ and $t_{f}$,
the latter probability tends to zero (so that $|F|^{2}$
approaches the unity) and the vacua $|0,t_{0}\rangle$ and
$|0,t_{f}\rangle$ become indistinguishable. Hence, the vacuum tends
asymptotically to be stable.

Let us remark that, as an immediate consequence of Eq.
(\ref{vac-rel}), ``particles'' are created in pairs. Besides, it
should be emphasized that ``particle'' production is due to the fact
that $\hat{N}(t)$ does not commute with the reduced
Hamiltonian,
\begin{equation} \left[:\widehat{\bar{H}}_{r}: ,
\hat{N}(t)\right]=\sum_{m=1}^{\infty}
\rho_{(m,t)}\left(\hat{b}_{m}(t)\hat{b}_{-m}(t)-\hat{b}^{\dagger}_{m}
(t)\hat{b}^{\dagger}_{-m}(t)\right). \end{equation} However, since
$\rho_{(m,t)}=1/(8mt^{2})$, the commutator vanishes in the
asymptotic limit of large times when the system becomes massless,
i.e. when the time dependent potential equals zero. So, the theory
becomes ``free" and the quantum representation approaches (the
counterpart of) the Poincar\'e-invariant one.

\section{Conclusion and further comments}
\label{sec:8}
\setcounter{equation}{0}

Let us summarize our results, discuss some consequences of the
introduced quantization and compare it with previous ones. The first
observation that is worth emphasizing is that, even if the field
$\hat{\xi}(t_{0};\theta)$ evolves unitarily to
$\hat{\xi}(t;\theta)$, the (explicitly time dependent) formal
operator $\hat{\phi}(t;\theta):=\hat{\xi}(t;\theta)/\sqrt{t}$, which
was regarded as the basic field for the quantum model in Ref.
\cite{pierri}, does not display a unitary evolution. In other words,
$\hat{\phi}(t_0;\theta)$ and $\hat{\phi}(t;\theta)$ are not related
by means of a unitary operator. The choice of fundamental field
plays therefore a decisive role in the construction of a
satisfactory quantization (together with the subsequent choice of
annihilation and creation operators, i.e. the complex structure
$J$). To arrive to the new field-parametrization, we have benefitted
from the freedom available to introduce a time dependent canonical
transformation and redistribute in that way the time dependence in
an implicit part, whose evolution is generated by the corresponding
reduced Hamiltonian of the model, and an explicit part (the factor
$1/\sqrt{t}$ for $\phi$ in our case), whose variation does not
necessarily have to be described by a unitary transformation. We
notice that, in systems like the symmetry reduced Gowdy model where
the Hamiltonian depends explicitly on time, it is natural to take
into account the possibility of performing canonical transformations
that vary with time. It is also worth pointing out that this system
exhibits what seems to be a general feature of quantum field
systems, in the sense that generic linear canonical transformations
do not become unitarily implemented in the quantum theory (see for
instance \cite{shale} and \cite{ccq:cqg}).

On the other hand, we have seen that the vacuum of the quantum
theory proposed for $\xi$ is not left invariant by the time
evolution. As a consequence, there is some ``particle" production by
the vacuum, which certainly attenuates as $t$ becomes large, but
never becomes strictly zero {\it except} in the limit in which the
time dependent potential vanishes, i.e. at infinitely large times.
However, we have seen in detail that this poses no problem for
unitarity and that the quantum theory is perfectly consistent, with
a well defined evolution that is compatible with the standard
probabilistic interpretation of quantum mechanics.

The fact that particles can be created (in pairs) in expanding
universes was realized already in the late sixties
\cite{parker-prl}. Since then, particle creation has been
extensively discussed and studied in diverse contexts in QFT (see
e.g. \cite{berger1,parker-2,many-many}), leading to remarkable
results as the so-called Hawking \cite{hawking} and Unruh
\cite{unruh} effects. In general, these results rest on the analysis
of the (specific) Bogoliubov transformation that relates the
canonical operators between the {\em in} and {\em out} states. In
the Gowdy $T^{3}$ cosmological model, as we have seen, the positive
and negative frequency parts of the basic field $\xi$ become mixed
during evolution owing to the time dependent  potential
$V(\xi)=\xi^{2}/(4t^{2})$. The canonical annihilation and creation
operators associated with {\em out} Fock states, at time
$T_{\rm{out}}$, are linear combinations of those associated with
{\em in} Fock states, at an earlier time $T_{\rm{in}}$. From the
unitary implementability of the dynamics, it follows that every {\em
in} state with a finite number of ``particles'' evolves to an {\em
out} state which also has a finite (although possibly different)
number of ``particles''.

This result applies even when $T_{\rm{out}}$ tends to infinity. In
that limit, a neat particle interpretation is available for the {\em
out} Fock states. Indeed, finiteness in the number of ``particles''
is a simple consequence of the fact that
$\lim_{T_{\rm{out}}\to\infty}|\beta_{n}(T_{\rm{out}},T_{\rm{in}})|
=|d_{n}(T_{\rm{in}})|$ and that $d_{n}(t)$ is square summable for
all $t>0$. On the other hand, the normalized {\em{out}} modes
$G^{(\infty)}_{n}(t,\theta):=\lim_{T_{\rm{out}}\to\infty}
\exp\left[i\delta_{|n|}T_{\rm{out}}-i\pi/4\right]
G^{(T_{\rm{out}})}_{n}
(t,\theta)$ (see Sec.~\ref{sec:7}) behave like the positive
frequency modes of Minkowski spacetime (except for the different
background topology) in the limit where the system becomes free.
Therefore, states in the {\em out} Fock space ${\cal{F}}_{\rm
{out}}({\cal{H}}_{\infty})$, which is constructed from the space of
solutions and the natural complex structure defined by the modes
$G^{(\infty)}_{n}$, admit a natural particle interpretation in the
asymptotic future. That is, ${\cal{F}}_{\rm
{out}}({\cal{H}}_{\infty})$ can be asymptotically identified with
the standard Fock representation of a free massless scalar field
propagating in a Minkowski spacetime, so that well defined
asymptotic notions of vacuum and particles arise. Clearly, in the
limit where the system is invariant under time translations,
ambiguities in the particle interpretation are avoided. But,
furthermore, an approximate adiabatic notion of particles
\cite{birrell-davies} can be introduced for each finite,
sufficiently large value of $T$, because in the asymptotic regime
the potential $V$ varies then very slowly in time. Thus, for large
$T_{\rm{in}}$ and $T_{\rm{out}}$, a notion of particles with a
conventional interpretation is available. One should,  of course,
keep in mind that the notion of ``particle'' we refer to is in the
sense of QFT on curved space, where the existence of a well defined
notion of particles refers to actual particles (as registered in
detectors). In the present quantum gravity system, even when its
degrees of freedom are captured by the scalar field, the
``particles'' associated with this field are far from having a clear
interpretation in terms of geometrical objects or ``quanta''.

The Gowdy model was reduced in Ref. \cite{pierri} to a free
massless scalar field $\phi$ on a flat, but time dependent
background, subject to a global constraint. We know that a
conventional quantization of this field does not allow one to
represent the evolution by means of a unitary transformation. By a
field redefinition, which involves the time parameter, we have
mapped the system into a scalar field $\xi$ satisfying a
``Klein-Gordon'' type equation in a background which is flat and
time independent (like 3-dimensional Minkowski spacetime, apart from
the topology), although in presence of a time dependent potential.
The natural quantization of this new field that we have presented in
this work provides a theory with a unitary dynamics. The relation
between this theory and the natural {\it time-translation invariant}
quantization that is available in this background {(namely, the
analog of the Poincar\'e invariant quantization in the considered}
spacetime) becomes manifest in the asymptotic regime where the
time-translation symmetry is recovered and a preferred vacuum with
that symmetry can be selected. More precisely, for asymptotically
large values of $T$, we have shown that our complex structure
$J_{T}$ approaches the ``Poincar\'e-invariant'' one in the limit in
which the system becomes massless and hence invariant under
time-translations. Although this result suggests the appealing
possibility that there exists a connection between unitary
implementability and asymptotic symmetries, further research is
needed to elucidate what kind of physical requirements lead to
acceptable quantizations (an equivalent of the Hadamard condition in
QFT in curved space \cite{wald}).

In order to arrive at a quantum description of the linearly
polarized Gowdy model one needs inputs at both the classical and
quantum levels. Classically, there are two important inputs. One
is the choice of deparametrization, namely the choice of a
fictitious time. The other is the field-parametrization adopted
for the metric. We have seen that, once a choice of time gauge has
been selected, the freedom in the choice of basic field can be
understood as that in performing time dependent canonical
transformations in the system. Besides, the order in which the two
previously mentioned choices are made in our case is irrelevant,
because it does not affect the final outcome, as we show in the
Appendix. In the quantum part, on the other hand, the cornerstone
is the choice of a complex structure (still a classical construct
on phase space), that determines the vacuum and the structure of
the quantum theory. In this respect, our choice of complex
structure $J$ was in some sense natural, guided both by previous
proposals for the quantization \cite{pierri} and by the symmetries
of the system in the asymptotic region of large times. It would be
very interesting to determine all other possible quantum
representations of the canonical commutation relations for our
specific field-parametrization (and time gauge) that permit a
unitary implementation of the dynamics, together with some
additional physical requirements, and elucidate whether they are
all unitarily equivalent. If this were so, the quantum theory
presented here would be essentially unique, once the choice of
internal time and fundamental field has been fixed. This issue
will be the subject of a future investigation \cite{ccmv}.

Finally, let us present some general comments on the validity of our
results in the more general context of quantum gravity. Quantum
gravity, both in its full glory and in reduced (midi-superspace)
models, suffers from the celebrated problem of time. Roughly
speaking, this means that there is no fundamental notion of time
(even classically) and that this has implications for the usual
probabilistic interpretation in the quantum theory. The Gowdy models
are not free from such a problem. There are basically two different
approaches, resulting from the two different ways of quantizing
constrained systems: quantize first and then reduce (Dirac) or
reduce first an then quantize (reduced phase space). The procedure
that has been followed here and in Refs.
\cite{pierri,ccq-t3,torre-prd} for the Gowdy model, although closer
to the second option, is a mixture of both approaches: one reduces
the system classically via gauge fixing and deparametrization, but
keeps a global constraint, which is dealt with in the quantum
theory. For the problem of time, one is choosing an internal time
$t$, via the deparametrization procedure, that takes the role of an
``external parameter'' in ordinary quantum theory. Of course, the
parameter $t$ is not the physical time even classically, but
provides us with the familiar framework of quantum theory to answer
real physical questions.

From the viewpoint of canonical quantum theory, where the theory is
defined over an abstract 3-manifold $\Sigma$, the natural picture
for the quantum description of gravity is the Heisenberg picture. A
quantum state $|\Psi\rangle$ of the system is defined on $\Sigma$
(that should {\it not} be thought of as a ``constant-time slice''
since there is no time and no spacetime to embed it), and
observables are of the Heisenberg type, namely evolving constants of
motion \cite{rovelli1,ccq-t3}. To be precise, we have for the Gowdy
model a family of operators $\{\hat{O}_i(t)\}$, one observable for
each value of $t$. One can, of course, define these operators and
relate them by means of the ``evolution operator'' $U(t,t_0)$ which
is, as we have shown, perfectly well defined. The Heisenberg picture
has to be contrasted with the Schr\"odinger one that in full quantum
gravity (in the Dirac approach) is simply not defined, since there
is no notion of embedding of the hypersurface $\Sigma$ on a
spacetime, much less the notion of ``time evolution''. In our case,
however, since we have defined a notion of (internal) time, we are
free to try and construct both the Schr\"odinger and the Heisenberg
pictures. As we have shown, these two pictures are well defined in
our model. Finally, let us end this note by pointing out that in
order to make full justice to the quantum geometry description given
by our choice of quantization, one would need to analyze the
behavior of (quantum) metric objects that provide a description of
the quantum geometry as in Ref.~\cite{pierri}, and the observables
recently introduced in Ref.~\cite{torre:obs}. We shall leave that
analysis for future research.

\section*{Acknowledgments}

The authors are greatly thankful to J.M. Velhinho for enlightening
conversations and fruitful interchange of ideas. This work was
supported by the Spanish MEC Projects No. FIS2004-01912 and No.
FIS2005-05736-C03-02 and by CONACyT (M\'exico) U47857-F grant.
J. Cortez was funded by the Spanish MEC, No./Ref. SB2003-0168.

\appendix
\section{Gauge fixing in the new field-parametrization}
\setcounter{equation}{0} \label{app-gt3-new}
\renewcommand{\theequation}{A.\arabic{equation}}

In this appendix, we explicitly show that the canonical
transformation (\ref{cano-transf}) amounts to a
field-reparametrization of the metric of the Gowdy model which
commutes with the process of gauge fixing \cite{jg}. In particular,
the gauge choice is not modified.

Let us start with the $3+1$ decomposition of the metric for the
polarized Gowdy $T^3$ spacetimes after fixing the gauge
corresponding to diffeomorphisms in the direction of the coordinates
$\sigma\in S^1$ and $\delta\in S^1$, with the two axial Killing
vector fields identified with $\partial_{\sigma}$ and
$\partial_{\delta}$ \cite{jg}:
\begin{equation} \label{metric-1-red} ds^2 = -N^2 dt^2 + h_{\theta
\theta}\big[d\theta + N^\theta dt\big]^2 +h_{\sigma\sigma}d\sigma^2
+h_{\delta\delta}d\delta^2.
\end{equation} Here, $N$ is the lapse function, $N^{\theta}$ is the
$\theta$-component of the shift vector, and $h_{ij}$ is the induced
spatial metric. To arrive at this expression, we have employed the
fact that the two Killing vector fields are hypersurface orthogonal,
so that $h_{\sigma\delta}$ must vanish. Besides, the presence of the
Killing symmetries implies that all metric functions depend only on
$\theta\in S^1$ and on the time coordinate $t$, which we choose
to be positive.

Instead of adopting the same field-parametrization as in Ref.
\cite{jg} for the induced metric, namely
\begin{equation}
\label{old-var} h_{\theta \theta}=e^{\gamma -\psi}, \qquad h_{\sigma
\sigma}=e^{- \psi}\tau^2 , \qquad h_{\delta \delta}=e^{\psi},
\end{equation} we now introduce an alternative parametrization in
terms of a new set of fields $\{Q^{\alpha}\}:=\{\xi , \bar{\gamma} ,
\tau\}$,
\begin{equation}
\label{new-var} h_{\theta \theta}=e^{\bar{\gamma}
-(\xi/\sqrt{\tau})- \xi^{2}/(4\tau)}, \qquad h_{\sigma \sigma}=
e^{-\xi/\sqrt{\tau}}\tau^2 , \qquad h_{\delta \delta}=e^{
\xi/\sqrt{\tau}}. \end{equation} With this new
field-parametrization, we obtain
\begin{equation} \label{metric-new-var-1} ds^2=e^{\bar{\gamma}
-(\xi/\sqrt{\tau})-\xi^{2}/(4\tau)}\left(-\tau^2
{N_{_{_{\!\!\!\!\!\!\sim}}\;}}^{2} dt^2+[d\theta +
N^{\theta}dt]^2\right) + e^{-\xi/\sqrt{\tau}}\left(\tau^2 d\sigma^2
+ e^{2\xi/\sqrt{\tau}}d\delta^2\right), \end{equation} where
${N_{_{_{\!\!\!\!\!\!\sim}}\;}}:=N/\sqrt{h}$ is the densitized lapse
function and $h$ is the determinant of the induced metric.

Regarding the change from the components of $h_{ij}$ to the set
$\{Q^{\alpha}\}$ as a point transformation, it is straightforward to
find momenta $P_{\alpha}$ canonically conjugate to $Q^{\alpha}$ in
terms of those for the induced metric \cite{jg}. In this way, one
arrives at the following Einstein-Hilbert action in Hamiltonian
form:
\begin{equation} \label{action-v1}
S = \int_{t_{i}}^{t_{f}}dt\oint d\theta \biggl[
P_{\tau}\dot{\tau}+P_{\bar{\gamma}}\dot{\bar{\gamma}}+P_{\xi}
\dot{\xi}-({N_{_{_{\!\!\!\!\!\!\sim}}\;}} \tilde{{\cal{C}}}+
N^{\theta}{\cal{C}}_{\theta})\biggr],
\end{equation} where the momentum and (densitized) Hamiltonian
constraints adopt the respective expressions \begin{eqnarray}
\label{const-mom}
{\cal{C}}_{\theta}&=&P_{\tau}\tau^{\prime}+P_{\bar{\gamma}}
\bar{\gamma}^{\prime}+ P_{\xi}\xi^{\prime}-
2P_{\bar{\gamma}}^{\prime},\\ \label{const-hamil}
\tilde{{\cal{C}}}&=&\frac{\tau}{2}P_{\xi}^2+
\frac{\xi^2}{8\tau}P_{\bar{\gamma}}^2 -\tau
P_{\tau}P_{\bar{\gamma}}+ \frac{\tau}{2}\left[ 4\tau'' -2
\bar{\gamma}^{\prime}\tau^{\prime} -\left(\frac{\xi
\tau^{\prime}}{2\tau}\right)^2 + (\xi^{\prime})^2\right].
\end{eqnarray}

On the other hand, a comparison between the field-parametrizations
(\ref{old-var}) and (\ref{new-var}) shows
\begin{equation}\label{chanfi}\gamma=\bar{\gamma}-
\frac{\xi^{2}}{4\tau},\qquad
\psi=\frac{\xi}{\sqrt{\tau}},\end{equation} with the same field
$\tau$ in both cases. This point transformation leads then to the
following relations between the corresponding canonical momenta:
\begin{eqnarray} \label{trans-psi} P_{\xi} = \frac{P_{\psi}}
{\sqrt{\tau}}-\frac{P_{\gamma}\psi}{2\sqrt{\tau}}, \qquad
P_{\tau}=\tilde{P}_{\tau}+\frac{P_{\gamma}\psi^2}{4\tau}-
\frac{P_{\psi}\psi}{2\tau} , \qquad P_{\bar{\gamma}}=P_{\gamma},
\end{eqnarray} where we have called $\tilde{P}_{\tau}$
the momentum conjugate to $\tau$ in the old parametrization, to
distinguish it from the new one, $P_{\tau}$.

In order to deparametrize the model and fix (almost all of) the
remaining gauge freedom, we must impose additional conditions that,
together with the constraints (\ref{const-mom}) and
(\ref{const-hamil}), form a set of second class constraints allowing
the reduction of the system. In Ref. \cite{jg}, the conditions
imposed were $g_{1}:=P_{\gamma}+p=0$ and $g_2:=\tau-tp=0$, where
$p=-\oint d\theta P_{\gamma}/(2\pi)$ is a constant of motion for the
model. Note that the only canonical variables that appear in these
conditions are $P_{\gamma}$ and $\tau$. With our change of metric
fields, $\tau$ is not modified and $P_{\gamma}$ becomes
$P_{\bar{\gamma}}$. Therefore, the gauge fixing selected to arrive
at the usual description of the Gowdy model, and that we choose to
impose also in the new field-parametrization, is
\begin{equation} \label{gauge-cond}
g_{1}:=P_{\bar{\gamma}}+p=0, \qquad g_{2}=\tau -tp=0.
\end{equation}

An analysis along the lines discussed in Ref. \cite{jg} shows that
the gauge fixing is well posed provided that $p\neq 0$. As explained
in the main text, we restrict all considerations to the sector of
positive $p$. The compatibility of the gauge fixing with the
dynamics sets ${N_{_{_{\!\!\!\!\!\!\sim}}\;}}=1/(pt)$, whereas
$N^{\theta}$ can be any function of $t$. Although the shift vector
is not entirely determined, the allowed functional form is such that
its contribution to the metric can be absorbed by means of a
redefinition of the coordinate $\theta$.

The momentum constraint ${\cal C}_{\theta}=0$ together with the
gauge fixing conditions imply that
\begin{equation} p\bar{\gamma}^{\prime}=P_{\xi}\xi^{\prime}.
\label{globalconst}
\end{equation} Since $p>0$, this relation determines the function
$\bar{\gamma}$, except for its zero mode. Given the periodicity of
the system in $\theta$, Eq. (\ref{globalconst}) also supplies the
homogenous constraint that remains on the system,
\begin{equation} \label{global-const}
C_{0}:=\frac{1}{\sqrt{2\pi}}\oint d\theta P_{\xi} \xi ^{\prime}=0.
\end{equation}

As a result of the commented gauge fixing, one finally obtains a
reduced system with spacetime metric
\begin{eqnarray} \label{metric-po-gow} ds^2&=&e^{\bar{\gamma}
-(\xi/\sqrt{pt})-\xi^{2}/(4pt)}
\left(-dt^{2}+d\theta^{2}\right)+e^{-\xi/\sqrt{pt}}
t^{2}p^{2}d\sigma^{2}+e^{\xi/\sqrt{pt}}d\delta^{2}, \\
\bar{\gamma}&=&-\frac{\bar{Q}}{2\pi p}-i\sum_{n=-\infty, n \neq
0}^{\infty}\frac{1} {2\pi np}\oint d\bar{\theta}e^{i n
(\theta-\bar{\theta})}P_{\xi} \xi^{\prime}+\frac{t}{4\pi p}\oint d
\bar{\theta}\left[P^{2}_{\xi}+(\xi^{\prime})^{2}+
\frac{1}{4t^{2}}\xi^{2}\right],
\end{eqnarray} where $\bar{Q}$ is the configuration variable
canonical to $\bar{P}=\ln{p}\in \mathbb{R}$. This metric coincides
in fact with that obtained from Eq. (\ref{metri}) (expressed in
terms of the field $\phi$) when the time dependent canonical
transformation (\ref{cano-transf}) is applied directly in the
reduced model.

In addition, the action that one obtains after the gauge
fixing procedure is (modulo a spurious boundary term)
\begin{equation}
\label{red-action-v1} S_{r}=  \int_{t_{i}}^{t_{f}}dt \left\{ \bar{P}
\dot{\bar{Q}} + \oint d \theta\left[P_{\xi}\dot{\xi}-
\frac{1}{2}\left(P^{2}_{\xi}+(\xi^{\prime})^{2}+\frac{1}{4t^{2}}
\xi^{2}\right)\right] \right\}.
\end{equation}
So, the reduced Hamiltonian is precisely that deduced in Eq.
(\ref{hamil-new-var}). Therefore, as we wanted to show, the
field-reparametrization commutes with the gauge fixing and reduction
procedure.


\begin{thebibliography}{99}

\bibitem{misner1} C.W. Misner, ``Minisuperspace'',
in {\em Magic without Magic: John Archibald Wheeler}, edited by J.
Klauder (Freeman, San Francisco, 1972).

\bibitem{torre} C.G. Torre,
``Midisuperspace Models of Canonical Quantum Gravity'', Int. J.
Theor. Phys. {\bf 38}, 1081 (1999), gr-qc/9806122.

\bibitem{gowdy} R.H. Gowdy, ``Vacuum Space-Times with two Parameter
Spacelike Isometry Groups and Compact Invariant Hypersurfaces:
Topologies and Boundary Conditions'', Ann. Phys. {\bf 83}, 203
(1974).

\bibitem{varios} C.W. Misner, ``A Minisuperspace
Example: The Gowdy $T^3$ Cosmology'', Phys. Rev. D {\bf 8}, 3271
(1973); B.K. Berger, ``Quantum Cosmology: Exact Solution for the
Gowdy $T^3$ Model'', Phys. Rev. D {\bf 11}, 2770 (1975).

\bibitem{berger1} B.K. Berger, ``Quantum Graviton Creation in
a Model Universe'', Ann. Phys. {\bf 83}, 458 (1974).

\bibitem{berger} B.K. Berger, ``Quantum Effects in the Gowdy $T^3$
Cosmology'', Ann. Phys. {\bf 156}, 155 (1984).

\bibitem{husain-smolin} V. Husain and L. Smolin, ``Exactly Solvable
Quantum Cosmologies from two Killing Field Reductions of General
Relativity'', Nucl. Phys. B {\bf 327}, 205 (1989).

\bibitem{guillermo-gowdy} G.A. Mena Marug\'an, ``Canonical
Quantization of the Gowdy model'', Phys. Rev. D {\bf{56}}, 908
(1997), gr-qc/9704041.

\bibitem{pierri} M. Pierri, ``Probing Quantum General Relativity
through Exactly Soluble Midi-Superspaces. II: Polarized Gowdy
Models'', Int. J. Mod. Phys. D {\bf 11}, 135 (2002), gr-qc/0101013.

\bibitem{ccq-t3} A. Corichi, J. Cortez, and H. Quevedo, ``On Unitary
Time Evolution in Gowdy $T^3$ Cosmologies'', Int. J. Mod. Phys. D
{\bf 11}, 1451 (2002), gr-qc/0204053.

\bibitem{torre-prd} C.G. Torre, ``Quantum Dynamics of the Polarized
Gowdy $T^3$ Model'', Phys. Rev. D {\bf 66}, 084017 (2002),
gr-qc/0206083.

\bibitem{jg} J. Cortez and G.A. Mena Marug\'an, ``Feasibility of a
Unitary Quantum Dynamics in the Gowdy $T^3$ Cosmological Model'',
Phys. Rev. D {\bf 72}, 064020 (2005), gr-qc/0507139.

\bibitem{jacob} T. Jacobson, ``Unitarity, Causality and Quantum
Gravity'', in {\em Conceptual Problems of Quantum Gravity}, edited
by A. Ashtekar and J. Stachel (Birkh\"auser, Boston, 1991).

\bibitem{heis-schr} P.A.M. Dirac, {\em Lectures on Quantum Field
Theory} (Yeshiva Univ., New York, 1966); ``Foundations of Quantum
Mechanics'', Nature (London) {\bf 204}, 115 (1964).

\bibitem{SandG} J.M. Simon and J.G. Taylor, ``Existence of the
Schr\"odinger and Heiseberg Pictures'', Nature (London) {\bf 205},
1305 (1965).

\bibitem{rovelli1} C. Rovelli,
``What is Observable in Classical and Quantum Gravity?'', Class.
Quantum Grav. {\bf 8}, 297 (1991); ``Time in Quantum Gravity: An
Hypothesis'', Phys. Rev. D {\bf 43}, 442 (1991).

\bibitem{ccmm} A. Corichi, J. Cortez, and G.A. Mena Marug\'an,
``Unitary Evolution in Gowdy Cosmology'',  Phys. Rev. D {\bf 73},
041502(R) (2006), gr-qc/0510109.

\bibitem{guillermo-varios} G.A. Mena Marug\'an and M. Montejo,
``Quantization of Pure Gravitational Plane Waves'', Phys. Rev. D
{\bf 58}, 104017 (1998), gr-qc/9806105; G.A. Mena Marug\'an, ``Gauge
Fixing and the Hamiltonian for Cylindrical Spacetimes'', Phys. Rev.
D {\bf 63}, 024005 (2001), gr-qc/0011068.

\bibitem{abramowitz} M. Abramowitz and I.A. Stegun (eds.),
{\em Handbook of Mathematical Functions} 9th edn. (Nat. Bur. Stand.
Appl. Math. Ser. No. 55) (Washington DC: US Government, 1970).

\bibitem{shale} D. Shale, ``Linear Symmetries of Free Boson
Fields'', Trans. Am. Math. Soc. {\bf 103}, 149 (1962).

\bibitem{honegger} R. Honegger and A. Rieckers,
``Squeezing Bogoliubov Transformations on the Infinite
CCR-Algebra'', J. Math. Phys. {\bf 37}, 4292 (1996).

\bibitem{ash-ma} A. Ashtekar and A. Magnon-Ashtekar,
``A Curiosity Concerning the Role of Coherent States in Quantum
Field Theory'', Pramana {\bf 15}, 107 (1980); ``A Geometrical
Approach to External Potential Problems in Quantum Field Theory'',
Gen. Rel. Grav. {\bf 12}, 205 (1980).

\bibitem{torre-vara-fevol} C.G. Torre and M. Varadarajan,
``Functional Evolution of Free Quantum Fields'', Class.  Quantum
Grav. {\bf 16}, 2651 (1999), hep-th/9811222.

\bibitem{wald} R.M. Wald, {\em Quantum Field Theory in Curved
Spacetime and Black Hole Thermodynamics} (Chicago Press, Chicago,
1994).

\bibitem{parker-2} L. Parker,
``Quantized Fields and Particle Creation in Expanding Universes.
1'', Phys. Rev. {\bf 183}, 1057 (1969).

\bibitem{ccq:cqg} A. Corichi, J. Cortez, and H. Quevedo,
``Note on Canonical Quantization and Unitary Equivalence in Field
Theory'', Class. Quantum Grav. {\bf 20}, L83 (2003), gr-qc/0212023.

\bibitem{parker-prl} L. Parker, ``Particle Creation in Expanding
Universes'', Phys. Rev. Lett. {\bf 21}, 562 (1968).

\bibitem{many-many} Y.B. Zel'dovich and A.A. Starobinsky,
``Particle Production and Vacuum Polarization in an Anisotropic
Gravitational Field'', Zh. Eksp. Teor. Fiz. {\bf 61}, 2161 (1971)
[Sov. Phys. JETP {\bf 34}, 1159 (1972)]; W.G. Unruh, ``Second
Quantization in the Kerr Metric'', Phys. Rev. D {\bf 10}, 3194
(1974); ``Origin of the Particles in Black Hole Evaporation'', Phys.
Rev. D {\bf 15}, 365 (1977); R.M. Wald, ``Stimulated Emission
Effects in Particle Creation near Black Holes'', Phys. Rev. D {\bf
13}, 3176 (1976); S.A. Fulling, ``Remarks on Positive Frequency and
Hamiltonians in Expanding Universes'', Gen. Rel. Grav. {\bf 10}, 807
(1979); G.T. Horowitz and R.M. Wald, ``Quantum Stress Energy in
Nearly Conformally Flat Space-Times'', Phys. Rev. D {\bf 21}, 1462
(1980); L. Parker, ``Cosmological Constant and Absence of Particle
Creation'', Phys. Rev. Lett. {\bf 50}, 1009 (1983); L.H. Ford,
``Gravitational Particle Creation and Inflation'', Phys. Rev. D {\bf
35}, 2955 (1987); J.A. Frieman, ``Particle Creation in Inhomogeneous
Space-Times'', Phys. Rev. D {\bf 39}, 389 (1989); B.L. Hu, G. Kang,
and A. Matacz, ``Squeezed Vacua and the Quantum Statistics of
Cosmological Particle Creation'', Int. J. Mod. Phys. A {\bf 9}, 991
(1994).

\bibitem{hawking} S.W. Hawking, ``Particle Creation by Black
Holes'', Commun. Math. Phys. {\bf 43}, 199 (1975).

\bibitem{unruh} W.G. Unruh, ``Notes on Black-Hole Evaporation'',
Phys. Rev. D {\bf 14}, 870 (1976).

\bibitem{birrell-davies} N. Birrell and P. Davies,
{\em Quantum Fields in Curved Space} (Cambridge University Press,
Cambridge, UK, 1982).

\bibitem{ccmv} A. Corichi, J. Cortez, G.A. Mena Marug\'an, and J. M.
Velhinho, {\it in preparation}.

\bibitem{torre:obs} C.G. Torre,
``Observables for the polarized Gowdy model'', Class. Quantum Grav.
{\bf 23}, 1543 (2006), gr-qc/0508008.

\end{thebibliography}
\end{document}